\documentclass[sigconf]{acmart} 

\usepackage{enumitem}
\usepackage{graphicx}
\usepackage{caption,subcaption}
\usepackage{color,colortbl}
\usepackage{array}

\usepackage{fancyhdr}
\usepackage{ctable}

\usepackage{url}
\usepackage{multirow}
\usepackage{booktabs} 
\usepackage{amsmath}
\usepackage{makecell}
\usepackage{cellspace}

\definecolor{Gray}{gray}{0.9}
\newcolumntype{M}[1]{>{\arraybackslash}m{#1}}

\def\BibTeX{{\rm B\kern-.05em{\sc i\kern-.025em b}\kern-.08emT\kern-.1667em\lower.7ex\hbox{E}\kern-.125emX}}

\copyrightyear{2020}
\acmYear{2020}
\setcopyright{iw3c2w3}
\acmConference[WWW '20]{Proceedings of The Web Conference 2020}{April 20--24, 2020}{Taipei, Taiwan}
\acmBooktitle{Proceedings of The Web Conference 2020 (WWW '20), April 20--24, 2020, Taipei, Taiwan}
\acmPrice{}
\acmDOI{10.1145/3366423.3380224}
\acmISBN{978-1-4503-7023-3/20/04}

\begin{document}

\title{Ten Social Dimensions of Conversations and Relationships}

\author{Minje Choi}
\affiliation{
  \institution{University of Michigan}
  \streetaddress{500 S State St}
  \city{Ann Arbor}
  \country{Michigan, US}}
\email{minje@umich.edu}

\author{Luca Maria Aiello}
\orcid{0000-0002-0654-2527}
\affiliation{
  \institution{Nokia Bell Labs}
  \streetaddress{21 JJ Thomson Avenue}
  \city{Cambridge}
  \country{United Kingdom}}
\email{luca.aiello@nokia-bell-labs.com}

\author{Kriszti\'an Zsolt Varga}
\affiliation{
  \institution{Nokia Bell Labs}
  \streetaddress{21 JJ Thomson Avenue}
  \city{Budapest}
  \country{Hungary}}
\email{krisztian.varga@nokia-bell-labs.com}

\author{Daniele Quercia}
\affiliation{
  \institution{Nokia Bell Labs}
  \streetaddress{21 JJ Thomson Avenue}
  \city{Cambridge}
  \country{United Kingdom}}
\email{quercia@cantab.net}

\begin{abstract}
Decades of social science research identified ten fundamental dimensions that provide the conceptual building blocks to describe the nature of human relationships. Yet, it is not clear to what extent these concepts are expressed in everyday language and what role they have in shaping observable dynamics of social interactions. After annotating conversational text through crowdsourcing, we trained NLP tools to detect the presence of these types of interaction from conversations, and applied them to 160M messages written by geo-referenced Reddit users, 290k emails from the Enron corpus and 300k lines of dialogue from movie scripts. We show that social dimensions can be predicted purely from conversations with an AUC up to 0.98, and that the combination of the predicted dimensions suggests both the types of relationships people entertain (conflict vs. support) and the types of real-world communities (wealthy vs. deprived) they shape.
\end{abstract}

%
%


\maketitle

\section{Introduction}\label{sec:intro}

Research in the social sciences dedicated considerable efforts to draw systematic categorizations of the fundamental sociological dimensions that describe human relationships~\cite{fiske1992four,wellman90different,bicchieri2005grammar,spencer2006rethinking}. This was partly motivated by the necessity to model relationships beyond tie strength~\cite{dedeo2013collective,aiello2014reading,Aiello2017}, as ties with equal strength may result into a wide variety of relationship types~\cite{marsden1984measuring,white2008notes,chowdhary2015ties}. Recently, such extensive literary production was surveyed by Deri et al.~\cite{deri18coloring}, who compiled an extensive review of decades' worth of findings in sociology and social psychology to identify \emph{ten dimensions} that have been widely used as ways to categorize relationships: \emph{knowledge}, \emph{power}, \emph{status}, \emph{trust}, \emph{support}, \emph{romance}, \emph{similarity}, \emph{identity}, \emph{fun}, and \emph{conflict} (description in Table~\ref{table:literature_review}). Although these categories are not meant to cover exhaustively all possible social experiences, Deri et al. provided empirical evidence that most people are able to characterize the nature of their relationships using these ten concepts only. Through a small crowdsourcing experiment, they asked people to spell out keywords that described their social connections (Table~\ref{table:literature_review}) and found that all of them fitted into the ten dimensions.

{\def\arraystretch{1.5}
\begin{table*}[ht]

\centering
\footnotesize
\begin{tabular}{p{10mm} p{70mm} p{70mm} p{15mm}}
\textbf{Dimension} & \textbf{Description} & \textbf{Keywords} & \textbf{References} \\
\Xhline{2\arrayrulewidth}
Knowledge & Exchange of ideas or information; learning, teaching &  teaching, intelligence, competent, expertise, know-how, insight & \cite{fiske2007universal,levin2004strength} \\
Power & Having power over the behavior and outcomes of another & command, control, dominance, authority, pretentious, decisions & \cite{french1959bases,french1956formal,blau64exchange} \\
Status & Conferring status, appreciation, gratitude, or admiration upon another &  admiration, appreciation, praise, thankful, respect, honor & \cite{blau64exchange,emerson1976social} \\
Trust & Will of relying on the actions or judgments of another & trustworthy, honest, reliable, dependability, loyalty, faith &  \cite{luhmann1982trust,zaheer1998does}\\
Support & Giving emotional or practical aid and companionship & friendly, caring, cordial, sympathy, companionship, encouragement &  \cite{baumeister1995need,fiske2007universal,vaux1988social}\\
Romance & Intimacy among people with a sentimental or sexual relationship &  love, sexual, intimacy, partnership, affection, emotional, couple & \cite{buss2003evolution,buss1993sexual,emlen1977ecology} \\
Similarity & Shared interests, motivations or outlooks & alike, compatible, equal, congenial, affinity, agreement &   \cite{mcpherson2001birds,jackson2010social} \\
Identity & Shared sense of belonging to the same community or group & community, united, identity, cohesive, integrated & \cite{tajfel2010social,oakes1994stereotyping,cantor1979prototypes} \\
Fun & Experiencing leisure, laughter, and joy &  funny, humor, playful, comedy, cheer, enjoy, entertaining & \cite{radcliffe1940joking,billig2005laughter,argyle2013psychology}\\
Conflict & Contrast or diverging views &  hatred, mistrust, tense, disappointing, betrayal, hostile & \cite{berlyne1960conflict,tajfel1979integrative} \\
\Xhline{2\arrayrulewidth}
\end{tabular}
\caption{The ten social dimensions of relationships studied by decades of research in the social sciences. The keywords are the most popular terms used by people to describe those dimensions, according to Deri at al.~\cite{deri18coloring}'s survey.}
\label{table:literature_review}
\vspace{-15pt}
\end{table*}
}

By combining these ten fundamental blocks in opportune proportions, one can draw an accurate, explainable, and intuitive description of the nature of most relationships, as perceived by the people involved. However, although the ten dimensions provide a useful way to conceptualize relationships, it is not clear to what extent these concepts are expressed through language and what role they have in shaping observable dynamics of social interactions. The growing availability of online records of conversational traces provides an opportunity to mine linguistic patterns for markers of their presence. Past research in Web Mining and Natural Language Processing (NLP) studied aspects pertaining some of the dimensions we deal with in this work~\cite{danescu2012echoes,ma2017self}, with special attention to concepts at the extremes of the spectrum of sentiment such as conflict~\cite{kumar18community} or empathy~\cite{morelli2017empathy,polignano2017learning} and support~\cite{wang18support,yang2019channel}. The operationalization of some of these concepts proved useful to improve the accuracy of prediction tasks~\cite{buntain2014identifying,wang2016learning,mitra2014language,wen2019finding}.

So far, little work has been conducted to explore all the ten dimensions systematically and jointly in relation to the use of language. In this study, we show that all ten social dimensions can be predicted purely from \emph{conversations}, and that the combination of the predicted dimensions suggests both the types of \emph{relationships} people entertain and the types of real-world \emph{communities} they shape. Specifically, we made three main contributions:
\begin{itemize}[leftmargin=*]
	\item We collected conversation records from various sources (\S\ref{sec:data}), and we labeled them according to the ten dimensions using crowdsourcing. We obtained annotations for a total of $\sim$9k texts and $\sim$5k Twitter relationships (\S\ref{sec:methods:crowdsourcing}), and found that all dimensions are abundantly expressed in everyday language (\S\ref{sec:results:crowdsourcing}).
	\item Using the collected data, we train multiple classifiers to predict the 10 dimensions purely from text (\S\ref{sec:methods:training}). Some dimensions are harder to predict because of their more complex lexical variations. Deep learning classifiers are more capable of handling such complexity, yielding an average AUC of 0.85 across the dimensions and a maximum AUC of 0.98 (\S\ref{sec:results:classification}). The model shows a good level of robustness when tested on unseen data sources.
	\item We find that the combination of the dimensions predicted from two individuals' conversations on Twitter predicts their type of social relationships (\S\ref{sec:crossdomain}). Further, by applying our framework to 160M messages written by geo-referenced Reddit users, 290k emails from the Enron corpus, and 300k lines of dialogue from movie scripts, we show that the presence of the ten dimensions in the language is indicative of the types of communities people shape (\S\ref{sec:qualitative}). For example, some of the dimensions are predictive of societal outcomes in US States, such as education, wealth, and suicide rates (\S\ref{sec:geostudy}). 
\end{itemize}

\section{Data collection}\label{sec:data}

To test our method on a diverse range of data, we extracted information about conversations and relationships from four sources.

\subsection{Reddit comments}\label{sec:dataset:reddit}

Reddit is a public discussion website, is one of the most accessed websites in the World and mostly popular in the United States where half of its user traffic is generated~\cite{alexa19reddit}. Reddit is structured in 140k+ independent \emph{subreddits} dedicated to a broad range of topics~\cite{medvedev2017anatomy}. Users can post a new \emph{submission} to a subreddit and write \emph{comments} to existing submissions. A dataset containing the vast majority of the submissions and comments published on Reddit since 2007 is publicly available~\cite{baumgartner15reddit,pushshift_data}. We gathered the data for the year 2017, which is nearly complete, according to recent estimates~\cite{gaffney2018caveat}. In total, we collected 96,212,869 submissions and 886,886,260 comments from 13,874,369 users.

To match Reddit discussions with census data (\S\ref{sec:geostudy}), we focused our analysis on users whom we could geo-reference at the level of US States. Reddit does not provide explicit information about user location, yet it is possible to get reliable location estimates with simple heuristics. Following the approach by Balsamo et al.~\cite{balsamo2019firsthand}, we first selected 2,844 subreddits related to cities or states in the United States~\cite{us_states_reddit}. From each of those, we listed the users who posted at least 5 submissions or comments. From the resulting set of users, we removed those who contributed to subreddits in multiple states. This resulted in 967,942 users who are likely to be located in one of the 50 US states. The number of users per state ranges from 1,042 (South Dakota) to 75,548 (California). In 2017, these users posted 9,553,410 submissions and 148,114,859 comments overall. We used this data to conduct a spatial analysis of the use of language (\S\ref{sec:geostudy}) and we sample from it to build our training set (\S\ref{sec:methods:crowdsourcing}).

\subsection{Enron emails}\label{sec:dataset:enron}

Enron Corporation was an American company founded in 1985 that went bankrupt in 2001, when its systematic practices of accounting fraud were exposed to the public. After the scandal and the resulting investigation, The Enron Email Dataset~\cite{klimt2004enron} was released to the public~\cite{enron_data} and became a popular resource for research in network science and Natural Language Processing~\cite{coffee2001understanding,klimt2004enron,diesner2005exploration,peterson2011email}. Messages include the full text and the email header. By filtering on the ``from:'' and ``to:'' fields, we obtained a corpus of 287,098 messages exchanged among 9,706 employees between year 2000 and 2001. In this study, we use a sample of annotated Enron emails to test our classifier's performance (\S\ref{sec:crossdomain}), and we look at its entirety to conduct a descriptive study (\S\ref{sec:qualitative}).

\subsection{Movie dialogs}\label{sec:dataset:movies}

Scripted movie dialogs are fictional yet plausible representations of conversations that span a wide spectrum of human emotions and relationship types. The Cornell Movie-Dialogs Corpus~\cite{danescu2011chameleons} is one of the most comprehensive open collections of movie scripts, containing 304,713 utterances exchanged between 10,292 pairs of characters from 617 movies. Past research used it to investigate the relationship between language and social interaction dynamics~\cite{danescu2012echoes}. We used it to test our classification system (\S\ref{sec:crossdomain}), and for conducting a qualitative analysis of its output (\S\ref{sec:qualitative}).

\subsection{Twitter relationships}\label{sec:dataset:twitter}

\begin{figure}[t!]
    \centering
		\includegraphics[width=0.75\linewidth]{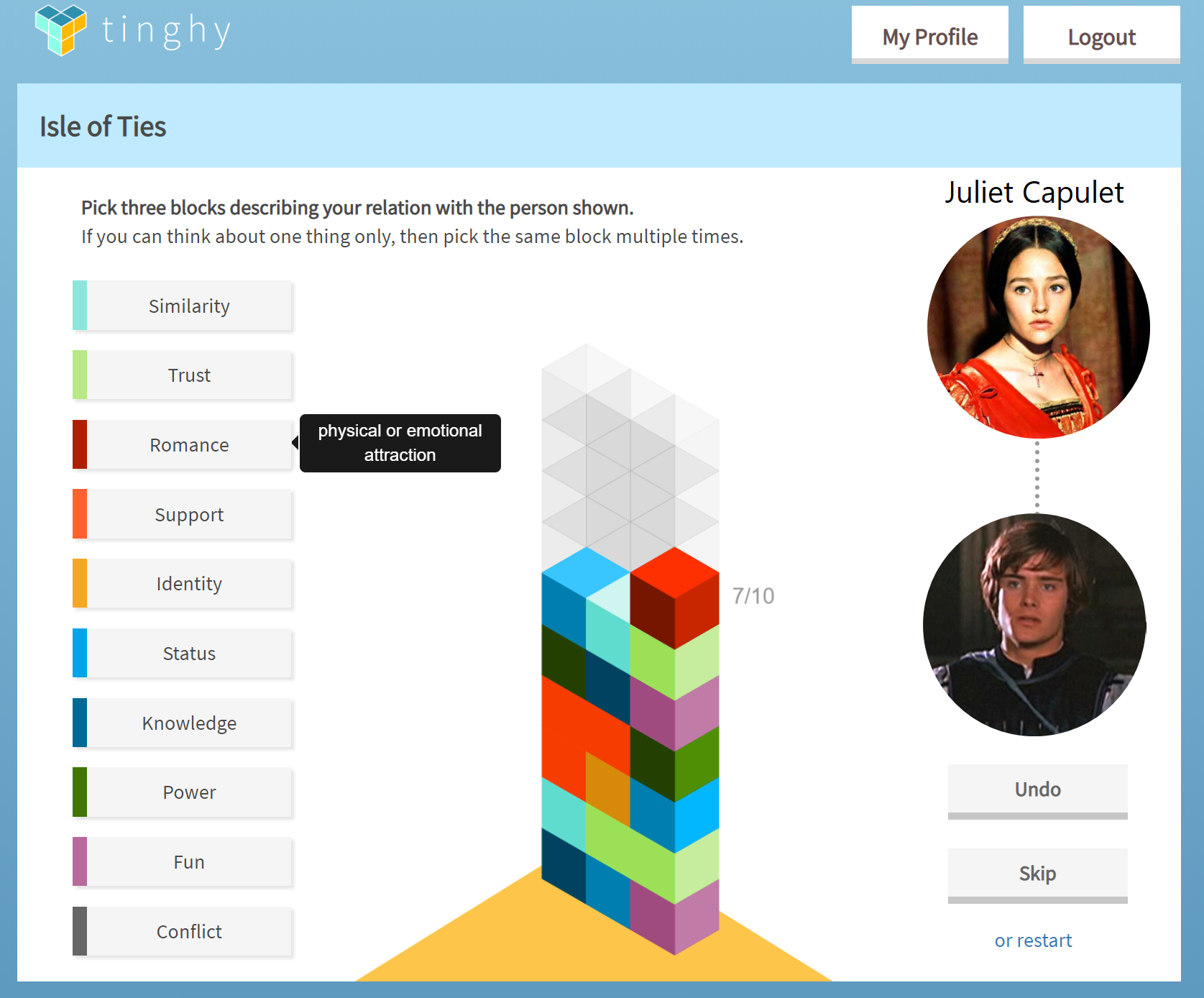}
    \caption{Anonymized screenshot of the Tinghy game. The player (bottom profile picture) is presented with 10 Twitter friends, one at the time (top profile picture) and is asked to describe their relationship by picking 1 to 3 dimensions from the menu on the left. By doing so, new blocks are added to the ``friendship wall'' in the middle. The dimensions are explained to the player with short text snippets.}
    \label{fig:figureeight_tinghy}
		\vspace{-10pt}
\end{figure}

Tinghy.org is a website that hosts a series of ``gamified'' psychological tests. Launched in 2018, it was conceived by Deri et al.~\cite{deri18coloring} as a platform to collect data about how social media users perceive their online relationships in terms of the 10-dimensional model of relationships. In one of these games, users log in with their Twitter account and they are sequentially presented with 10 of their Twitter followees. The selection of contacts is biased towards the strongest ties with the player. This is done using a validated linear regression model (see Table 1 in~\cite{gilbert2012predicting}) that estimates tie strength through a number of factors that can be calculated from the data exposed by the public Twitter API (e.g., time elapsed since last interaction). The player picks one to three dimensions over the 10 available to describe their relationship with each of the friends displayed (Figure~\ref{fig:figureeight_tinghy}). With the explicit user consent, interaction data is gathered through the Twitter API. For every player-friend pair ($u,v$), the dataset contains \emph{i)} a list of up to three dimensions picked by $u$, sorted by order of selection; \emph{ii)} the list of all tweets in which $u$ mentions (or replies to) $v$ or viceversa; and \emph{iii)} the list of $u$'s tweets that were retweeted by $v$ or viceversa. To date, 684 people played the game, providing labels for 5,217 social ties between a total of 3,777 unique individuals (the data was recorded even when players quit the game before completion). These ties exchanged 9,960 mentions, 31,100 replies and 8,619 retweets overall. We restricted our study to English tweets that account for 1,772 relationships between 1,406 unique individuals for a total of 8,870 mentions, 19,254 replies and 5,050 retweets.

Unlike the ground-truth labels for the other datsets, which are at sentence-level (\S\ref{sec:methods:crowdsourcing}), the annotations coming from this game are provided at relationship-level. This allowed us to test the extent to which one could predict the dominant social dimension of a relationship from conversations (\S\ref{sec:crossdomain}).

\section{Methodology}\label{sec:methods}

We adopted a supervised approach to extract the ten social dimensions from text. We crowdsourced a dataset of conversational texts annotated with the 10 dimensions (\S\ref{sec:methods:crowdsourcing}), and we used it to train multiple classifiers (\S\ref{sec:methods:training}).

\subsection{Crowdsourcing}\label{sec:methods:crowdsourcing}

To annotate text, we followed the same procedure for Reddit comments, movie dialogs, and Enron emails. For each data source, we split all texts into sentences, and retain only the sentences that contain at least one $1^{st}$ or $2^{nd}$ person pronoun. This filtering step is meant to bias the selection in favor of phrases that follow a conversational structure. We then selected a random sample of sentences with length between 6 and 20 words, to avoid statements that are too complex to assess or too short to be informative. For each sentence, we also kept the preceding and following sentences from the same text, if any. The addition of neighboring sentences is helpful for the annotators---albeit not strictly necessary---to make better sense of the context around the sentence.

\begin{figure}[t!]
    \centering
    \includegraphics[width=0.99\linewidth]{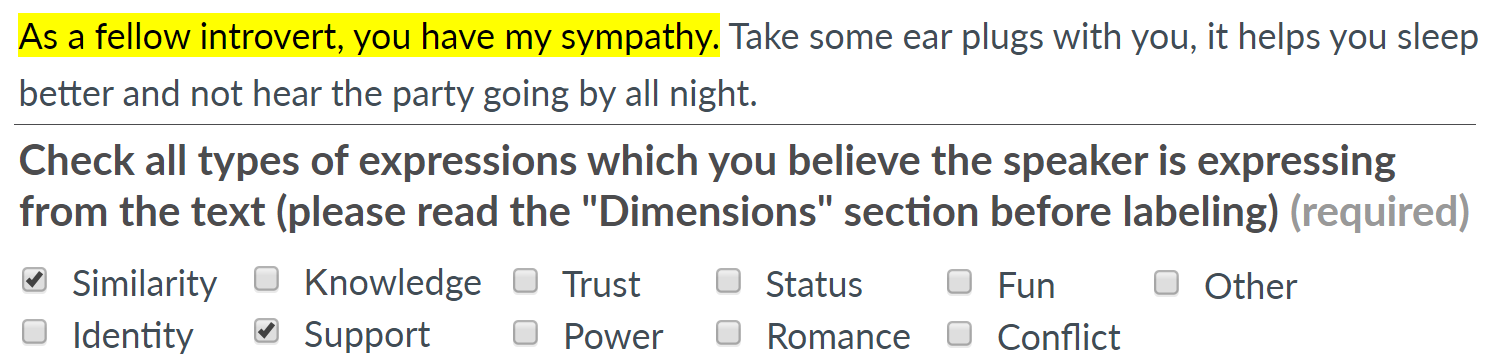}\\
		\caption{Example of the crowdsourcing task. The highlighted sentence conveys a combination of social support and similarity.}
    \label{fig:figureeight}
		\vspace{-10pt}
\end{figure}

Each resulting \emph{passage}, composed by the target sentence highlighted with color and surrounded by the neighboring phrases, is presented to crowdworkers for annotation. We asked them to read the whole passage and to select the dimensions that they believe the highlighted sentence conveys, among the 10 provided (Figure~\ref{fig:figureeight}). Annotators were encouraged to select multiple dimensions when they felt that more than one applied. A special label \emph{``other''} was provided in case the annotator was uncertain or no available option seemed pertinent. Each sentence was annotated by three people. 

Before starting the task, annotators read the definitions of all the 10 dimensions, which were extended versions of the statements in Table~\ref{table:literature_review}. For example, social support was described as: \emph{``Expressions that suggest the offer of any type of emotional or practical aid, which might come in different forms, including: sympathy, compassion, empathy, companionship, offering to help.''} Definitions were accompanied by 3 to 5 examples (e.g., for social support: \emph{``I am so sorry for your loss.''}). Instructions were accessible at any time during the task, for quick reference.

As a quality-control mechanism, we inserted \emph{test sentences} both at the beginning of each task and at random positions in the task. These consists of variations of the examples provided in the instructions, for which the correct dimension is known. The test sentences served two purposes. First, whenever an annotator provided a wrong answer to a test sentence, the correct answer was shown, so that they could learn from their mistakes. Second, annotators who failed to assign correct labels to $40\%$ of the test sentences or more were banned from the task, and their answers were discarded. Through small-scale preliminary tests, we empirically observed that 40\% was a good threshold to filter out misbehaving users.

We deployed the task on the crowdsourcing platform ``Figure Eight''. We opened the participation only to people residing in five English-speaking countries (United States, United Kingdom, Ireland, Canada, Australia) and who belong to the platform's top-tier expert contributors. We set the price for each annotation task to 0.05\$, which amounts to a 9\$ hourly wage considering an average time of 20 seconds spent on each sentence. We collected labels for 7,855 sentences from Reddit posts, 400 from movie lines, and 436 from Enron emails, which were provided by 934 annotators who labeled 28 sentences each on average. Workers spent $23s$ per sentence on average ($\sigma=35s$). The reported level of satisfaction after the task was 4.0 out of 5, on average.

\subsection{Classification}\label{sec:methods:training}

\subsubsection{Classifiers}\label{sec:methods:training:classifiers}

We experiment with four classification frameworks: a traditional ensemble classifier, a simple metric based on distance between words in an embedding space, and two deep-learning models.

\vspace{4pt}\noindent\textbf{Xgboost.} An ensemble of decision trees with gradient boosting~\cite{chen2016xgboost}. It is well-suited to small datasets, makes it easy to interpret the contribution of individual features, and is able to ignore any vacuous features that may be present to prevent overfitting. Xgboost has proven to be the best performing classifier among competitors in popular challenges. We train Xgboost using the features defined in \S\ref{sec:methods:training:features}, computed at sentence-level. We performed grid search to tune its learning rate and the maximum depth of its trees. In a binary classification task, Xgboost outputs a confidence score in [0,1] that captures the likelihood of the sample belonging to the positive class. 

\vspace{4pt}\noindent\textbf{Embedding distance.} Word embeddings are dense vector representations of words that capture the linguistic context in which words occur in a corpus. Such representations are generally learned by training neural network models on large text corpora to predict the occurrence of words from their local lexical context. Each word is associated with a point in the embedding space such that words that share common contexts are close to one another. Many embedding techniques have been developed recently~\cite{li2018word}, and several pre-trained models are readily available. GloVe~\cite{pennington2014glove} embeddings with 300 dimensions, trained on the Common Crawl corpus (42B tokens) performed best in the tasks we addressed. In addition to considering a word's local context, GloVe uses also global co-occurrence statistics across the whole text corpus.

We leveraged the properties of the embedding space to implement a simple measure of distance between a sentence and each of the 10 conversational dimensions. We first computed a sentence-level embedding vector $\mathbf{g}_s$ by averaging the embedding vectors of all the words in a sentence $s$:
\begin{equation}
\mathbf{g}_s = \frac{1}{len(s)} \cdot \sum_{w \in s} \mathbf{g}_w,
\label{eq:sentence_vector}
\end{equation}
where $\mathbf{g}_w$ is the GloVe vector of word $w$. We used the same formula to compute an embedding vector $\mathbf{g}_d$ for the words representative of each dimension $d$, as listed in Table~\ref{table:literature_review}. We then computed the Euclidean distance between the two resulting vectors: $d(\mathbf{g}_s, \mathbf{g}_d)$. This method yields a single measure that does not offer a natural threshold for binary classification, yet one that can rank sentences by their `relevance' to a dimension.

\vspace{4pt}\noindent\textbf{LSTM.} Long Short-Term Memory (LSTM)~\cite{hochreiter1997long} is a type of recurrent neural network (RNN) particularly suited to process data that is structured in temporal or logical sequences. LSTMs have demonstrated to achieve excellent results in timeseries forecasting~\cite{lipton2015critical,greff2016lstm} as well as in NLP tasks~\cite{sundermeyer2012lstm}. LSTM accepts fixed-size inputs; in our experiments, we fed it with a 300-dimension GloVe vector of a word, one word at a time for all the words in a sentence. Each new word updates the model's status by producing a new hidden-state vector. Following the standard approach, we applied a linear transformation to reduce the last hidden vector into one scalar value, and we apply a sigmoid function to transform it into a continuous value between 0 and 1, which indicates the probability of belonging to the positive class. We experimented with a simple LSTM model with no attention, short-cut connections, or other additions. We performed grid search to tune its hyperparameters (learning rate and number of epochs).

\vspace{4pt}\noindent\textbf{BERT.} Transformers~\cite{vaswani2017attention} are models designed to handle ordered sequences of data by relying on attention mechanisms rather than on recurrence. As opposed to directional models like LSTM, which read the input sequentially, transformers parse an entire sequence of words at once, thus allowing the model to learn the context of a word based on all of its surroundings (left and right context). BERT (Bidirectional Encoder Representations from Transformers) is a language representation model based on Transformers and pre-trained on a 3.3B word corpus from BooksCorpus and Wikipedia~\cite{devlin2018bert}. It has been adapted to solve several NLP tasks, achieving state-of-the-art results. We used a pretrained BERT-Base Cased model. Following the original specifications~\cite{devlin2018bert}, we fine-tune it to perform binary classification by adding a classification layer on top of the Transformer output, which results into a 2-dimensional output vector representing the two output classes. Last, we apply a softmax transformation to get a single score in [0,1] that reflects the likelihood of the input belonging to the positive class. We performed grid search to tune its learning rate and the number of epochs.

\subsubsection{Interpretable features}\label{sec:methods:training:features}

{\def\arraystretch{1.75}
\begin{table}[t!]
\footnotesize
\begin{tabular}{p{19mm} p{50mm} p{7mm}}
        \textbf{Feature family} & \textbf{Feature names} & \textbf{\# feat.}\\
    \toprule
        Linguistic style  & politeness~\cite{brown1987politeness,danescu2013computational}; hedging terms~\cite{fu2017confidence}; morality-related words~\cite{haidt2007morality}; integrative complexity~\cite{robertson2019language}; syntactic markers~\cite{tchokni2014emoticons}: word elongations, use of capital words, \#question marks, \#exclamation marks, \#ellipsis & 50 \\
				Readability \& complexity & \#words; avg. length of words; avg. syllable per word; entropy of words~\cite{tan2016winning}; readability indices~\cite{jurafsky2000speech}: Kincaid, ARI, Coleman-Liau, Flesch Reading Ease, Gunning-Fog index, SMOG index, Dale Challenge index & 12 \\
				Linguistic lexicons & LIWC~\cite{pennebaker2001linguistic}; Empath~\cite{fast2016empath} & 175 \\
        Sentiment  & VADER~\cite{hutto2014vader}; Hatesonar~\cite{davidson2017automated} & 6 \\
				Word distribution & n-grams~\cite{jurafsky2000speech} & 100\\
		\Xhline{2\arrayrulewidth}
    \end{tabular}
		\caption{Interpretable linguistic features for classification}
    \label{tab:feature_list}
		\vspace{-20pt}
\end{table}
}

To train the Xgboost model, we extracted a total of 343 features, partitioned in five families (Table~\ref{tab:feature_list}). We picked these sets of features because they have been successfully used to solve a variety of NLP tasks, they are intuitively interpretable, and they cover several facets of language use. Here we summarize them shortly and we refer the reader to the original publications for the detailed formulations. The first family of features captures aspects of \emph{linguistic style}: the use of formulas of politeness~\cite{danescu2013computational} and complex argumentation~\cite{fu2017confidence,robertson2019language}; the presence of words that appeal to morality~\cite{haidt2007morality}; and the use of a number of simple syntactic markers~\cite{tchokni2014emoticons}. The second one comprises a measures of \emph{readability and writing complexity}, ranging from simple counts to more sophisticated indices~\cite{jurafsky2000speech}. The third one includes LIWC~\cite{pennebaker2001linguistic} and Empath~\cite{fast2016empath}, two widely used \emph{linguistic lexicons} that map words into linguistic, psychological, and topical categories. The fourth one captures the spectrum of \emph{sentiment} with VADER~\cite{hutto2014vader}, a rule-based tool to measure positive/negative emotions in short text, and Hatesonar~\cite{davidson2017automated}, a tool to detect offensive language. Last, to capture the \emph{distribution of words}, we counted a sentence's unigrams and bigrams. To reduce the sparsity of the $n$-gram space, we considered only those that occur 10 times or more in the training set and we filtered them using log-odd Dirichlet priors to further narrow the set to those $n$-grams that are highly discriminative~\cite{monroe2008fightin}. Specifically, we kept only the top 100 $n$-grams ranked by $\xi = \log p(w|w \in P_{d}) - \log p(w)$, where $p(w)$ is the probability of a $n$-gram $w$ occurring in the full corpus, and $p(w|w \in P_{d})$ is the probability of occurring in the sentences of the positive set for the target dimension ($P_d$).

\subsubsection{Task definition}\label{sec:methods:training:setup}

Given a sentence $s$ and a social dimension $d \in D = \{d_1, \ldots, d_{10}\}$, our task was to determine whether $s$ conveys $d$. Rather than training one multi-class classifier, we treated each dimension independently and trained multiple binary classifiers. This choice was motivated by the non-exclusive nature of the ten dimensions~\cite{deri18coloring}: a sentence may convey any pair (or subsets) of dimensions at once---which we confirmed in our results (\S\ref{sec:qualitative}).

Given a dimension $d$, we included in its set of positive samples $P_d$ all the sentences that were labeled with $d$ by two annotators or more, and we put all the sentences never labeled with $d$ in the set of negative samples $N_d$. In each round of a 10-fold cross-validation, we randomly split each set in 80\% for training, 10\% for tuning, and 10\% for testing. Since $|P_d| < |N_d| \forall d$, we performed random oversampling~\cite{ling1998data} to balance the classes. Specifically, within each training, tuning, and testing split, we added multiple copies of positive samples picked at random until the size of the two classes got balanced. Compared to other oversampling techniques~\cite{chawla2002smote,he2008adasyn}, random oversampling does not generate synthetic data points, which might end up exhibiting unrealistic features. Its application is equivalent to giving higher importance to positive samples: classifying a positive instance correctly yields a performance gain that is proportional to the number of replicas (or an equally great loss if misclassified).

We measured performance with the average ``Area Under the ROC Curve'' across all folds---AUC, in short. AUC measures the ability of the model to correctly rank positive and negative samples by confidence score, independent of any fixed decision threshold. Because the data is balanced, the expected value of AUC for a random classification is 0.5.

\section{Results}\label{sec:results}

\subsection{Conversations}\label{sec:results:crowdsourcing}

Most agreement scores are well-defined for sets of items judged by all raters. We compute an inter-annotator agreement score on the set of test sentences which have been rated by all annotators. On this set, the Fuzzy Kappa agreement score~\cite{kirilenko2016inter}---an extension of Cohen's Kappa that contemplates the possibility of an instance being placed in multiple categories~\cite{mchugh2012interrater}---is 0.45, which indicates moderate agreement. On the full set, no consensus was reached on 41\% of the sentences, which were assigned no dimension. Some agreement is reached for the remaining 59\%: 53\% were assigned exactly one dimension, 5\% two, and 1\% three or more. Source-specific proportions are listed in Table~\ref{table:dimensions_distr}. Despite the selection of sentences was performed at random, almost 60\% of those from Reddit carry a social value that could be linked to the 10 dimensions. In movie scripts, this fraction raises to 90\%, which is expected considering that the narrative structure of movies compresses dense information about character relationships in a limited number of lines. Next, we focused on those sentences on which annotators reached some consensus, and used the remaining ones only as negative examples for training. In~\S\ref{sec:discussion}, we discuss the nature of the sentences for which no consensus was reached.

{\def\arraystretch{1.75}
\begin{table}[t]
\centering
\footnotesize
\begin{tabular}{p{6mm}p{6mm}p{2.5mm}p{2.5mm}p{2.5mm}p{2.5mm}}
& & \multicolumn{4}{c}{\textbf{\#Dimensions}}\\
\cline{3-6}
\textbf{Data} & \textbf{total\#} & 0 & 1 & 2 & 3+ \\
\Xhline{2\arrayrulewidth}
All    & 8,691 & 41\% & 53\% & 5\%  & 1\% \\
Reddit & 7,855 & 43\% & 54\% & 3\%  & 0\% \\
Movies & 400   & 10\% & 59\% & 24\% & 7\% \\
Enron  & 436   & 22\% & 59\% & 14\% & 5\% \\
\Xhline{2\arrayrulewidth}
\end{tabular}
\caption{Fraction of messages labeled with $n$ numbers of dimensions from the annotators}
\label{table:dimensions_distr}
\vspace{-10pt}
\end{table}
}

\begin{figure}
    \centering
    \includegraphics[width=0.99\linewidth]{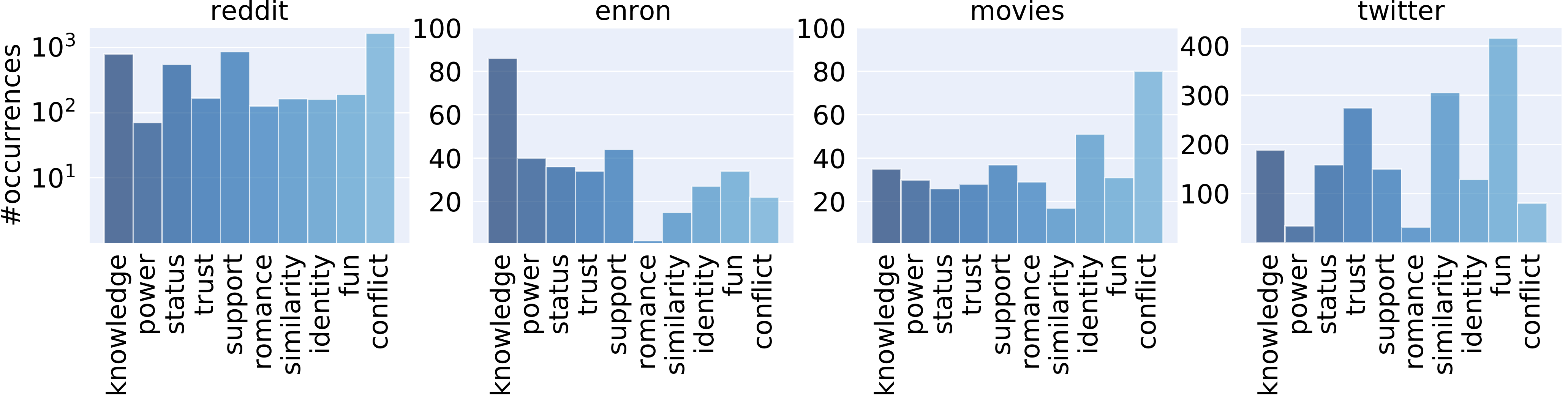}
    \caption{Distributions of labels across datasets.}
    \label{fig:distributions}
		\vspace{-10pt}
\end{figure}

Verbal expressions do not represent all dimensions in equal measure, and the relative proportions vary considerably across data sources (Figure~\ref{fig:distributions}). In Reddit, conflict is predominant, followed by support, knowledge, and status. This is in line with previous work that showed that Reddit communities are often aimed at providing social support~\cite{de2014mental,cunha2016effect,de2016discovering}, but they are also prone to fall prey of misbehaving users~\cite{cheng2017anyone,kumar18community}. In Enron, the relative abundance of knowledge-exchange messages reflects the nature of goal-oriented communication in corporations; unsurprisingly, romance is non-existent. Lines from movie scripts exhibit high level of conflict and identity, likely due to how fictional story arcs pivot around overcoming interpersonal challenges~\cite{field2005screenplay}, often instantiated by cohesive factions opposing each other~\cite{wolfenstein2002movie}. For Twitter relationships, the dominant dimensions are fun, similarity, trust, and knowledge, which reflect partly the bias of the data collection towards strong ties, and partly the nature of Twitter as a community of interest in which like-minded people exchange information~\cite{kwak2010twitter,conover2011political}.

\subsection{Classifying conversations}\label{sec:results:classification}

{\def\arraystretch{1.75}
\begin{table}[t]
\centering
\footnotesize
\begin{tabular}{l p{2.5mm}p{2.5mm}p{2.5mm}p{2.5mm}p{2.5mm}p{2.5mm}p{2.5mm}p{2.5mm}p{2.5mm}p{2.5mm}p{2.5mm}}
 & \multicolumn{8}{c}{\textbf{Xgboost}} &   &  \\
\cline{2-9}
 & \rotatebox[origin=l]{90}{Linguistic } & \rotatebox[origin=l]{90}{Hate } & \rotatebox[origin=l]{90}{Readability } & \rotatebox[origin=l]{90}{VADER } & \rotatebox[origin=l]{90}{Empath } & \rotatebox[origin=l]{90}{LIWC} & \rotatebox[origin=l]{90}{Ngrams } & \rotatebox[origin=l]{90}{All } & \rotatebox[origin=l]{90}{\textbf{Embedding}} & \rotatebox[origin=l]{90}{\textbf{LSTM}} & \rotatebox[origin=l]{90}{\textbf{BERT}} \\
\hline
\rowcolor{Gray} Knowledge & 0.61 & 0.6 & 0.65 & 0.66 & 0.7 & 0.77 & 0.69 & 0.76 & 0.53 & \textbf{0.82} & \textbf{0.82} \\
Power & 0.54 & 0.56 & 0.57 & 0.58 & 0.68 & <0.5 & 0.58 & 0.54 & 0.53 & \textbf{0.82} & 0.74 \\
\rowcolor{Gray}Status & 0.67 & 0.58 & 0.61 & 0.78 & 0.78 & 0.79 & 0.78 & 0.82 & 0.78 & \textbf{0.86} & 0.85 \\
Trust & 0.7 & <0.5 & 0.61 & 0.76 & 0.72 & 0.75 & 0.76 & \textbf{0.80} & 0.72 & 0.77 & 0.73 \\
\rowcolor{Gray}Support  & 0.62 & 0.55 & 0.64 & 0.69 & 0.75 & 0.78 & 0.69 & 0.79 & 0.66 & 0.83 & \textbf{0.85} \\
Romance & 0.85 & 0.53 & 0.77 & 0.82 & 0.97 & 0.93 & 0.82 & 0.96 & 0.78 & \textbf{0.98} & 0.93 \\
\rowcolor{Gray}Similarity & 0.5 & 0.53 & 0.55 & 0.62 & 0.63 & 0.6 & 0.62 & 0.63 & 0.64 & 0.80 & \textbf{0.82} \\
Identity & <0.5 & <0.5 & 0.57 & 0.50 & 0.55 & 0.67 & 0.62 & 0.59 & 0.66 & \textbf{0.75} & 0.62 \\
\rowcolor{Gray}Fun & <0.5 & 0.62 & 0.71 & 0.76 & 0.86 & 0.86 & 0.65 & 0.95 & 0.83 & 0.94 & \textbf{0.98} \\
Conflict & 0.57 & 0.57 & 0.64 & 0.79 & 0.75 & 0.81 & 0.61 & 0.84 & 0.66 & 0.86 & \textbf{0.91} \\
\Xhline{2\arrayrulewidth}
\end{tabular}
\caption{Performance of different models on each dimension for the Reddit dataset (average AUC over 10-fold cross validation). Top performances are highlighted in bold.}
\label{table:model_auc}
\vspace{-10pt}
\end{table}
}

Prediction results are summarized in Table~\ref{table:model_auc}. Among all the prediction models, the embedding similarity performed worst. LSTM and BERT reached comparable performances, yielding top scores on 5 dimensions each, with a tie on \emph{knowledge}; their performance gap is minor in most dimensions, with peak performances ranging from 0.75 to 0.98. AUC generally drops when using the Xgboost model, even when relying on all available features. Xgboost obtained the best performance on \emph{trust} only, and by a small margin.

\begin{figure}
    \centering
    \includegraphics[width=0.99\linewidth]{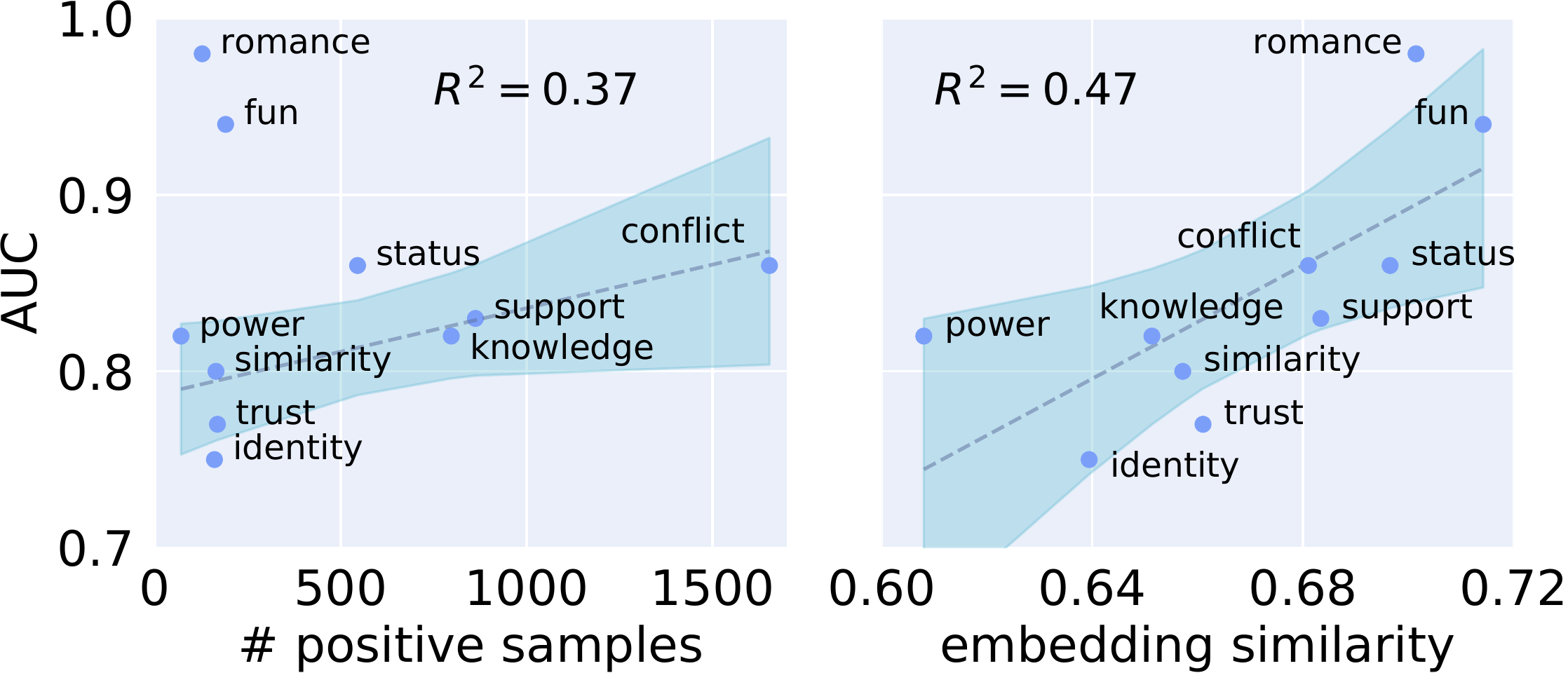}
    \caption{AUC increases with the size of the training data (left) and with the lexical homogeneity of the expressions used to express a dimension, estimated with average similarity in the embedding space (right).}
    \label{fig:auc_vs_similarity}
		\vspace{-10pt}
\end{figure}

Across classifiers, results suggest that some dimensions are easier to predict than others. For example, simple lexicons for sentiment analysis reach AUC scores exceeding 0.85 for \emph{fun} and \emph{romance}. To check the link between performance and size of training data, we plot the AUC against the number of positive samples for each dimension (Figure~\ref{fig:auc_vs_similarity}, left---LSTM only, for brevity). The AUC increases linearly with the dataset size ($R^2 = 0.37$) except when considering two outliers: \emph{romance} and \emph{fun}, which are associated with good performances despite the scarcity of their training data. We hypothesize that this discrepancy is due to the diverse nature of verbal expressions: the more limited the language variations used to express a dimension, the easier to predict those variations. To verify it, we computed the sentence-level embedding vectors (using Formula~\ref{eq:sentence_vector}) for all sentences in the sets of positive samples $P_d, \forall d$. We then measure the average cosine similarity between 100k random pairs of sentences within the same set $P_d$, which gives an estimate on how semantically close the verbal expressions in each dimensions are. We find a significant linear relationship ($R^2 = 0.47$) between average embedding similarity and AUC (Figure~\ref{fig:auc_vs_similarity}, right). As expected, \emph{romance} and \emph{fun} are the ones with highest similarity. This trend holds for all classifiers but it is particularly pronounced for Xgboost and for the simple embedding similarity baseline.

We conclude that, although Xgboost yields decent performances in some cases, its effectiveness suffers from the higher lexical variety of expressions in some dimensions (e.g., \emph{power} or \emph{identity}) more than that of deep learning models. Nevertheless, the nature of the Xgboost framework allows us to study the importance of its interpretable features in predicting different dimensions, thus providing a human-readable indication of whether the content of verbal exchanges in the labeled data matches theoretical expectations. We measure each feature's effect size using Cohen's $d$, and report only those with $d > 0.4$, which corresponds to a substantial effect size~\cite{cohen2013statistical}. Table~\ref{tab:important_features} shows the important features organized into each feature category. The features that emerge echo the theoretical definition of the ten dimensions (Table~\ref{table:literature_review}). Naturally, sentiment is an important feature for most. Pleasant interactions express positive sentiment, \emph{knowledge} and \emph{power} tend to be neutral, and \emph{conflict} carries negative sentiment. Furthermore: \emph{knowledge} is associated with complex writing; \emph{romance}, \emph{support}, and \emph{trust} with a sense of empathy and attachment; \emph{power} with work-related topics and with words conveying authority; \emph{fun} with words of play and celebration; \emph{similarity} with verbal formulas of comparison.

For simplicity, in the remainder of the paper we report only results for LSTM, which is computationally simpler and faster than BERT, and achieved similar results.

{\def\arraystretch{1.25}
\begin{table}[t!]
    \centering
		\footnotesize
    \begin{tabular}{p{12mm} p{63mm}}
        \textbf{Dimension} & \textbf{Top features} \\
        \toprule
				Knowledge & Readability (ARI, Kincaid, Gunning Fog Index, avg. words per sentence); VADER (neutral); Style (hedging)\\
				Power & Liwc (power, work); Vader(neutral); Empath (order, business, power)\\
				Status & Liwc (affect, posemo); Vader (positive); Empath (giving, optimism, politeness)\\
				Trust & Liwc (posemo, affect); Vader (positive); Empath (friends, help, trust); Style (empathy words) \\
				Support & Liwc (posemo); Vader (positive); Empath (optimism, help, giving); Ngram (``thank you''); Style (empathy words)\\
				Romance & Empath (affiliation, affection, friends, sexual, wedding, optimism); Style (empathy words); Liwc (affiliation, bio, social, drives, ppron, posemo) Vader (positive); Ngram (``love'')\\
				Similarity & Liwc (compare); Empath (appearance); Ngram (``like''); Style (integration words) \\
				Identity & Liwc (religion); Hatesonar (hatespeech); Empath (sexual)\\
				Fun & Empath (celebration, childish, children, fun, leisure, party, ridicule, toy, vacation, youth, optimism); Liwc (affect, posemo); Vader (positive); Style (``!'')\\
				Conflict &  Vader (negative); Liwc (anger, negate, swear, negemo); Readability (Dale Challenge Index); Empath (hate, swearing terms); Hatesonar (offensive language) \\
		\Xhline{2\arrayrulewidth}
    \end{tabular}
		\caption{Important feature groups per dimension in the Xgboost classifier (Cohen's $d > 0.4$)}
		\label{tab:important_features}
		\vspace{-10pt}
\end{table}
}

\subsection{Classifying relationships}\label{sec:crossdomain}

\begin{figure*}[t!]
    \centering
    \includegraphics[width=0.75\linewidth]{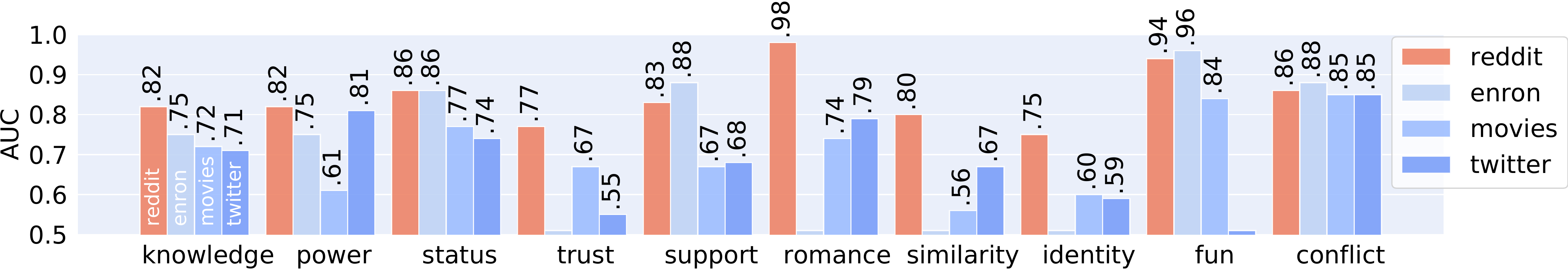}
		\hspace{15pt}
		\includegraphics[width=0.18\linewidth]{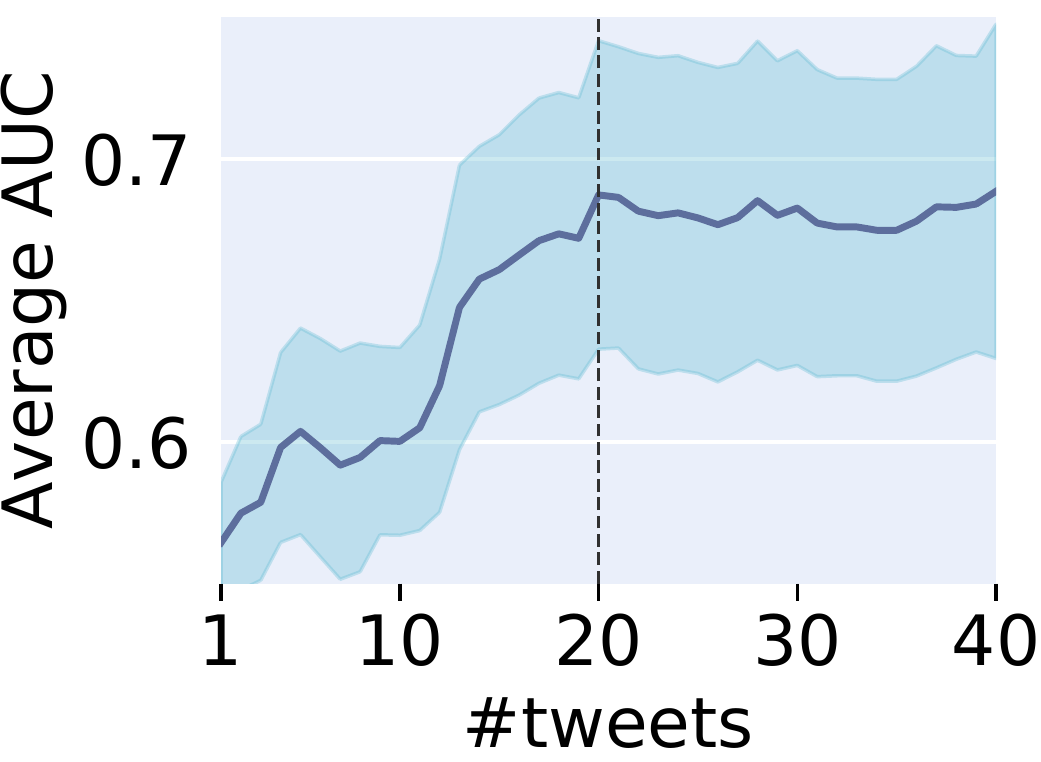}
    \caption{Left: AUC of LSTM models trained on the Reddit data and tested on the other datasets. Right: growth of AUC in the classification of Twitter relationships as the number of messages exchanged between the two users increases.}
    \label{fig:cross_dataset}
		\vspace{-10pt}
\end{figure*}

To test the adaptability of our model to different domains, we trained dimension-specific LSTM classifiers on all the available Reddit data and tested them on the corpora from Enron and movie scripts. Results are summarized in Figure~\ref{fig:cross_dataset} (left). 

In Enron, the performance did not drop when detecting \emph{status}, \emph{support}, \emph{fun} and \emph{conflict}, whereas \emph{knowledge} and \emph{power} suffered a loss within $0.1$. The AUC dropped when detecting utterances of \emph{similarity} and \emph{identity}, which both rarely appear in our labeled Enron sample. The model adapted to a lesser extent to movie scripts, arguably because the composition of scripted text is intrinsically different from user-generated text in blog posts or emails. Still, we recorded limited or no AUC loss for four dimensions out of ten (\emph{knowledge}, \emph{status}, \emph{fun}, and \emph{conflict}). As we shall see in our qualitative analysis (\S\ref{sec:qualitative}), even the lowest-performing classifiers dimensions returned meaningful results when applied to larger data sources and only high-confidence sentences were kept.

Last, we used the data collected from the Tinghy game to address an even more challenging task: predicting \emph{relationship-level} labels from conversations. For every pair of Twitter users $u, v$, we considered only the first dimension that $u$ picked in the game; the first association that comes to mind is likely to be the most relevant and important, according to several models of human attention~\cite{broadbent1957mechanical,fleming1998web,cutrell2007you}. We leave a multi-dimensional analysis of relationships to future work. We ran our classifier on the text of each mention, reply and retweet between the two users, disregarding the directionality of interaction. We estimated a relationship-level label by picking the most frequent dimension across all the messages.

We observed that the average AUC across dimensions grows with the volume of messages exchanged between the users. After a minimum of 20 messages, the performance reaches a plateau (Figure~\ref{fig:cross_dataset}, right). Therefore, we limited the prediction only to pairs of users who were involved in at least 20 interactions. In this setting, the prediction worked best (Figure~\ref{fig:cross_dataset}, left) for \emph{conflict} and \emph{status} ($AUC > 0.8$), and for \emph{power}, \emph{support}, and \emph{romance} ($AUC > 0.7$).

Overall, models that predict \emph{conflict}, \emph{status}, and \emph{knowledge} were the most robust across sources. Predictions suffered limited losses for about half of the dimensions in each dataset, which is remarkable given the limited size of training data. Finally, with the predictions on Twitter relationships, we produced evidence that the model could learn the perceived nature of a social tie from the conversations that flow over it.

\subsection{Qualifying conversations and relationships}\label{sec:qualitative}

We provide a qualitative assessment of the output of our tool on the Enron emails and on the movie scripts.

\subsubsection{The fall of Enron}

\begin{figure}[t!]
    \centering
    \includegraphics[width=0.99\linewidth]{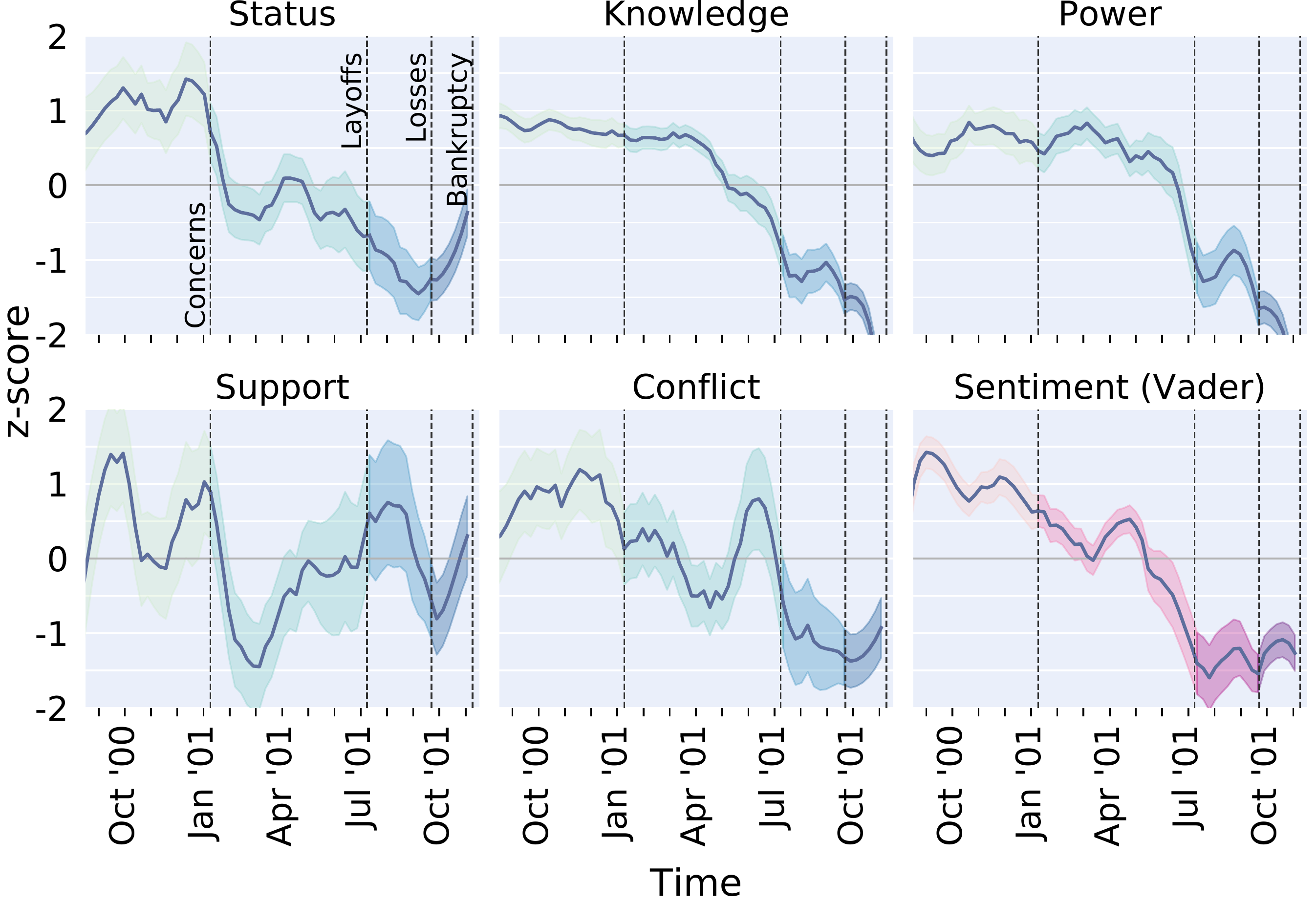}
    \caption{How the presence of five social dimensions in Enron employees emails changes over time, compared to a sentiment analysis baseline. Status giving, knowledge transfer, and the power-based exchanges plummet after the first financial concerns. After massive layoffs, the remaining employees give support to each other.}
    \label{fig:enron}
		\vspace{-10pt}
\end{figure}

The ability of identifying a rather comprehensive set of dimensions from conversational text enables us to interpret social phenomena with broader nuances compared to traditional tools like sentiment analysis. Both the longitudinal nature of the Enron dataset and the well-documented stages of the company's downfall make it possible to test whether exogenous events impact the presence of certain social dimensions in people's exchanges and relationships.

We ran our ten LSTM models $M_d, d \in \{1 \ldots 10\}$ on every email, and marked a text $t$ with dimension $d$ if the maximum confidence score for dimension $d$ across all its sentences is higher than $0.95$, namely $max(\{M_d(s), \forall s \in t\}) > 0.95$. In other words, a text conveys a dimension if at least one of its sentences is predicted with high confidence to express that dimension.  For all the emails sent during a calendar week $t$, we calculated the ratio $f_d(t)$ between emails carrying dimension $d$ and the total numbers of emails sent. Finally, we transformed these fractions into z-scores to make the values comparable across dimensions:
\begin{equation}
zscore_d(t) = \frac{f_d(t) - \mu_d}{\sigma_d}
\label{eqn:zscore_enron}
\end{equation}
where $\mu_d$ and $\sigma_d$ are the average and standard deviation of $f_d$ across all weeks. 

Figure~\ref{fig:enron} shows the trends of the dimensions over time. We excluded from the analysis those dimensions that did not perform well in the cross-domain adaptation of our models (Figure~\ref{fig:cross_dataset}). For the sake of comparison, we report also the z-score of the sentiment score calculated with VADER. All plots are marked with four significant events in Enron's history: \emph{i)} the beginning of widespread concerns about the financial stability of the company; \emph{ii)} the first round of layoffs; \emph{iii)} the start of financial losses; \emph{iv)} the declaration of bankruptcy. The picture traced by sentiment analysis marks an overall, steady downward trend that reaches its lowest level by the time financial losses were made official. The conversational dimensions, on the other hand, reveal a richer picture that matches the known stages of the company's downfall~\cite{mclean2013smartest}. First, as the initial concerns sparked, the exchange of status and support plummeted: panic started to spread and employees stopped celebrating their achievements, thanking each other, and offering comfort. About three months later, the frequency of knowledge exchange dropped sharply: as concerns grew, employees spent less time in dealing with their everyday duties. A few weeks before the layoffs, as it became clear that many employees would have been made redundant, conflict exploded and the power structure collapsed---fewer orders were given to the angry crowd of employees who were made aware of the impeding jobs cuts. In the aftermath of the layoffs, those who managed to stay in the company gave support to each other for a few weeks before the imminent crack.

\subsubsection{Movies}

Movie dialogues present dense and relatable narratives. Often the story and background of characters is laid out to the audience, which makes it easy to interpret their interactions. This motivated us to manually inspect some lines extracted by our machine learning tool. We ran our models on all lines from the movie script corpus, sorted them by confidence scores, and reported the top three for every dimension. In Table~\ref{tab:movie_quotes}, alongside each line, we report the histogram of confidence scores of the classifiers for all the dimensions. We observe that different dimensions can coexist and complement each other in various forms. For example, the sentence: \emph{``I want to thank you, sir, for giving me the opportunity to work''} (Table~\ref{tab:movie_quotes}, line 7) conveys \emph{status}, \emph{trust}, and \emph{support} at the same time (the speaker is thanking a respectable ``sir'' for trusting him with a job that will help him and his family out). Furthermore, the co-occurrence of dimensions shows how they could act as basic blocks that compose more complex sociological constructs. For example, utterances that combine power and knowledge express authoritativeness (Table~\ref{tab:movie_quotes}, lines 1,4), knowledge and identity may express cultural traditions (line 20), and the oscillation between power dynamics and trust is at the base of bargaining (line 5).

\begin{table*}
    \centering
		\footnotesize
    \begin{tabular}{M{1mm} M{1mm} M{138mm} M{26mm}}
				& &  & \raisebox{-\totalheight\relax}{\includegraphics[width=0.15\textwidth]{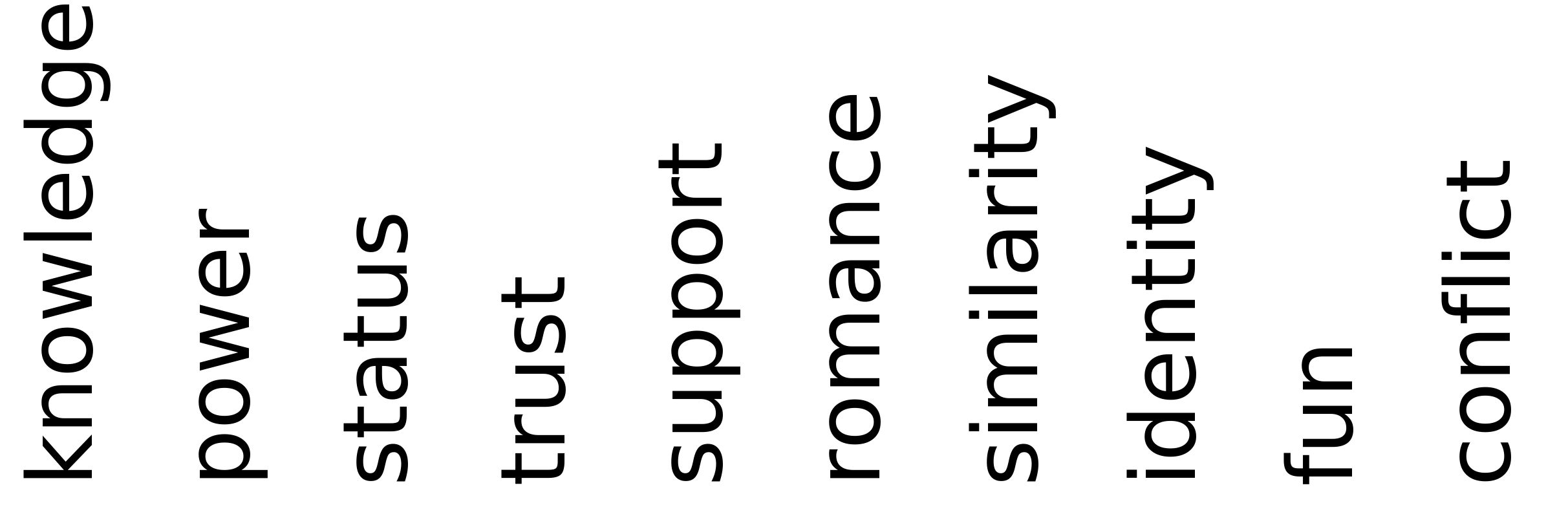}} \\
				\midrule
				\multirow{3}{*}[-2mm]{\rotatebox[origin=c]{90}{Knowledge}} & 1 &  \cellcolor{Gray} Only a fully trained Jedi Knight, with The Force as his ally, will conquer Vader and his Emperor. If you end your training now, if you choose the quick and easy path, as Vader did, you will become an agent of evil --- \textbf{Ben Kenobi, Star Wars Ep.5} &  \cellcolor{Gray} \raisebox{-\totalheight\relax}{\includegraphics[width=0.15\textwidth]{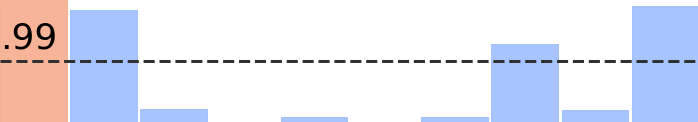}} \\ 
				& 2 & Well, in layman's terms, you use a rotating magnetic field to focus a narrow beam of gravitons; these in turn fold space-time consistent with Weyl tensor dynamics until the space-time curvature becomes infinitely large and you have a singularity --- \textbf{Dr. Weir, Event Horizon} &   \raisebox{-\totalheight}{\includegraphics[width=0.15\textwidth]{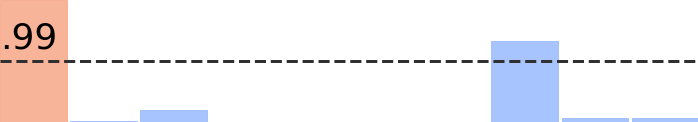}}\\ 
        & 3 & \cellcolor{Gray} Since positronic signatures have only been known to emanate from androids such as myself, it is logical to theorize that there is an android such as myself on Kolarus III --- \textbf{Data, Star Trek: Nemesis} &  \cellcolor{Gray} \raisebox{-\totalheight}{\includegraphics[width=0.15\textwidth]{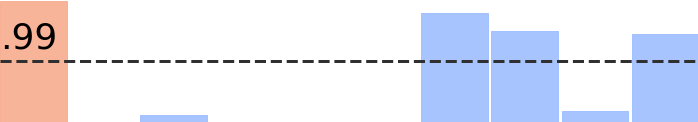}}\\ 
				\midrule 
        \multirow{3}{*}[-3mm]{\rotatebox[origin=c]{90}{Power}} & 4 &  Now if you don't want to be the fifth person ever to die in meta-shock from a planar rift, I suggest you get down behind that desk and don't move until we give you the signal --- \textbf{Ray Stantz, Ghostbusters II} &  \raisebox{-\totalheight\relax}{\includegraphics[width=0.15\textwidth]{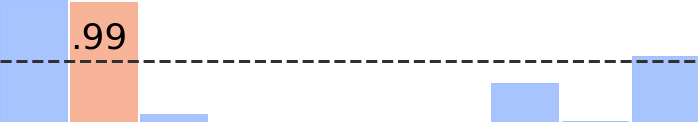}}  \\ 
				& 5 & \cellcolor{Gray} You can ask any price you want, but you must give me those letters --- \textbf{Ilsa Lund, Casablanca} & \cellcolor{Gray} \raisebox{-\totalheight\relax}{\includegraphics[width=0.15\textwidth]{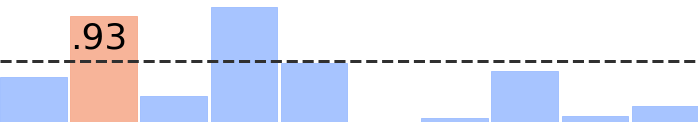}} \\ 
        & 6 & Right now you're in no position to ask questions! And your snide remarks... --- \textbf{Hunsecker, Sweet Smell of Success} &  \raisebox{-\totalheight\relax}{\includegraphics[width=0.15\textwidth]{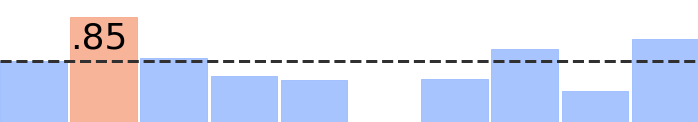}} \\ 
				\midrule 
				\multirow{3}{*}[-3mm]{\rotatebox[origin=c]{90}{Status}}  & 7 & \cellcolor{Gray} I want to thank you, sir, for giving me the opportunity to work --- \textbf{Mr. L\"ownstein, Schindler's List} & \cellcolor{Gray} \raisebox{-\totalheight\relax}{\includegraphics[width=0.15\textwidth]{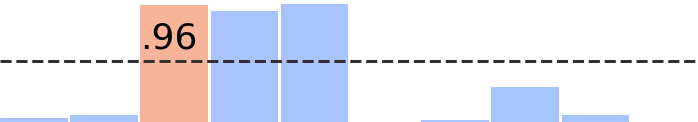}}\\
				& 8 & Frankie, you're a good old man, and you've been loyal to my Father for years...so I hope you can explain what you mean --- \textbf{Michael Corleone, The Godfather: Part II}  & \raisebox{-\totalheight\relax}{\includegraphics[width=0.15\textwidth]{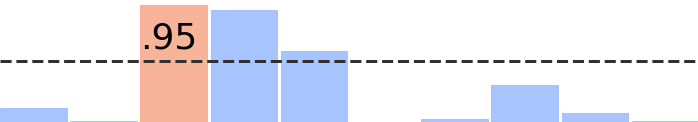}}\\ 
        & 9 &  \cellcolor{Gray} And we drink to her, and we all congratulate her on her wonderful accomplishment during this last year...her great success in A Doll's House! --- \textbf{Evan, Hannah and Her Sisters} & \cellcolor{Gray} \raisebox{-\totalheight\relax}{\includegraphics[width=0.15\textwidth]{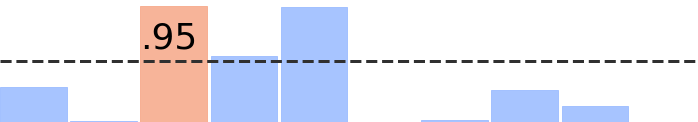}}\\ 
        \midrule 
        \multirow{3}{*}[-3.4mm]{\rotatebox[origin=c]{90}{Trust}} & 10 & I'm trying to tell you -- and this is where you have to trust me -- but, I think your life might be in real danger --- \textbf{Jack, Fight Club} & \raisebox{-\totalheight\relax}{\includegraphics[width=0.15\textwidth]{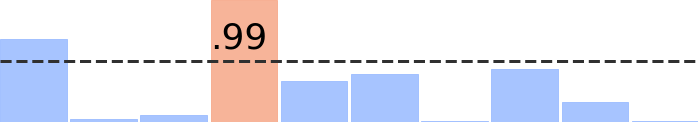}}\\
        & 11 &  \cellcolor{Gray} Mr. Lebowski is prepared to make a generous offer to you to act as courier once we get instructions for the money --- \textbf{Brandt, The Big Lebowski} & \cellcolor{Gray} \raisebox{-\totalheight\relax}{\includegraphics[width=0.15\textwidth]{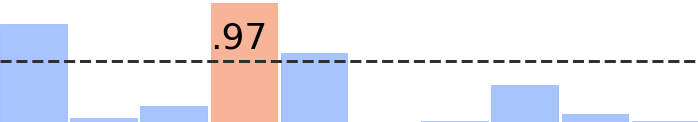}}\\ 
        & 12 & Take the Holy Gospels in your hand and swear to tell the whole truth concerning everything you will be asked --- \textbf{Pierre Cauchon, The Story of Joan of Arc} & \raisebox{-\totalheight\relax}{\includegraphics[width=0.15\textwidth]{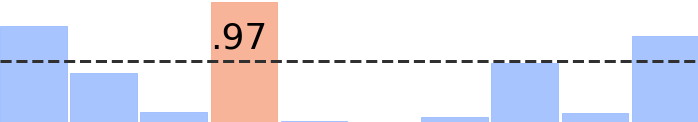}}\\
				\midrule 
				\multirow{3}{*}[-3.2mm]{\rotatebox[origin=c]{90}{Support}} & 13 & \cellcolor{Gray} I'm sorry, I just feel like... I know I shouldn't ask, I just need some kind of help, I just, I have a deadline tomorrow --- \textbf{Barton, Barton Fink} & \cellcolor{Gray} \raisebox{-\totalheight\relax}{\includegraphics[width=0.15\textwidth]{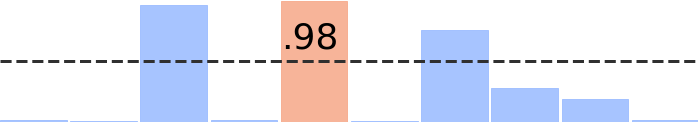}} \\ 
        & 14 &  Look, Dave, I know that you're sincere and that you're trying to do a competent job, and that you're trying to be helpful, but I can assure the problem is with the AO-units, and with your test gear --- \textbf{HAL 9000, 2001: A Space Odyssey} &  \raisebox{-\totalheight\relax}{\includegraphics[width=0.15\textwidth]{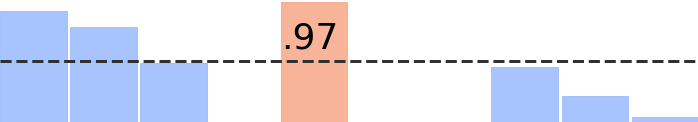}}\\ 
				& 15 & \cellcolor{Gray} Well... listen, if you need any help, you know, back up, call me, OK? --- \textbf{Detective Tania Johnson, Rush Hour}&  \cellcolor{Gray} \raisebox{-\totalheight\relax}{\includegraphics[width=0.15\textwidth]{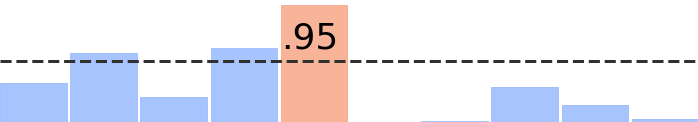}}\\
        \midrule 
				\multirow{3}{*}[-2.5mm]{\rotatebox[origin=c]{90}{Romance}} & 16 & I'm going to marry the woman I love --- \textbf{Harold, Harold and Maude} &  \raisebox{-\totalheight\relax}{\includegraphics[width=0.15\textwidth]{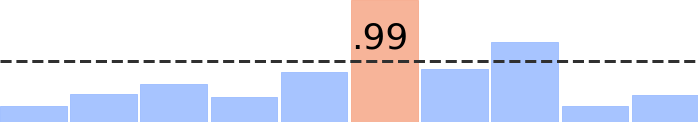}}\\
        & 17 & \cellcolor{Gray} If you are truly wild at heart, you'll fight for your dreams... Don't turn away from love, Sailor --- \textbf{The Good Witch, Wild at Heart} & \cellcolor{Gray} \raisebox{-\totalheight\relax}{\includegraphics[width=0.15\textwidth]{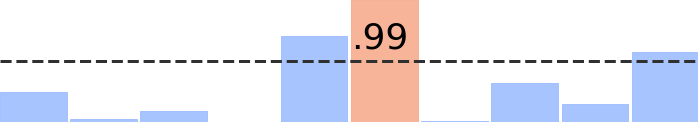}}\\
        & 18 &  You admit to me you do not love your fiance? --- \textbf{Westley, The Princess Bride}&  \raisebox{-\totalheight\relax}{\includegraphics[width=0.15\textwidth]{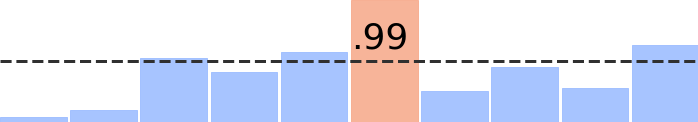}}\\
				\midrule 
				\multirow{3}{*}[-2.6mm]{\rotatebox[origin=c]{90}{Identity}} & 19 & \cellcolor{Gray} Hey, I know what I'm talkin' about, black women ain't the same as white women --- \textbf{Mr. Pink, Reservoir Dogs} & \cellcolor{Gray} \raisebox{-\totalheight\relax}{\includegraphics[width=0.15\textwidth]{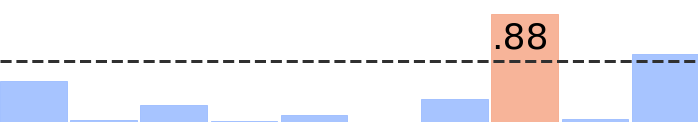}} \\
        & 20 & That's how it was in the old world, Pop, but this is not Sicily --- \textbf{Michael Corleone, The Godfather: Part II}& \raisebox{-\totalheight\relax}{\includegraphics[width=0.15\textwidth]{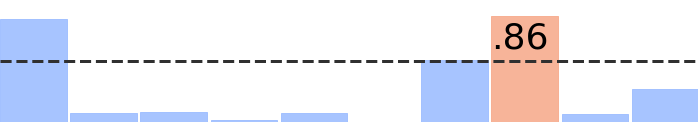}}\\ 
        & 21 & \cellcolor{Gray} But, as you are so fond of observing, Doctor, I'm not human --- \textbf{Spock, Star Trek: The Wrath of Khan}& \cellcolor{Gray} \raisebox{-\totalheight\relax}{\includegraphics[width=0.15\textwidth]{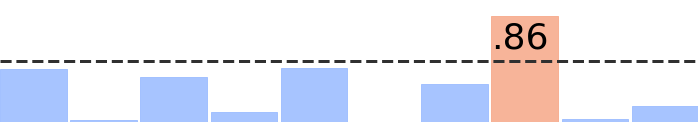}}\\ 
        \midrule 
				\multirow{3}{*}[-3.5mm]{\rotatebox[origin=c]{90}{Fun}} & 22 &  It's just funny...who needs a serial psycho in the woods with a chainsaw when we have ourselves --- \textbf{Pixel, Happy Campers} &  \raisebox{-\totalheight\relax}{\includegraphics[width=0.15\textwidth]{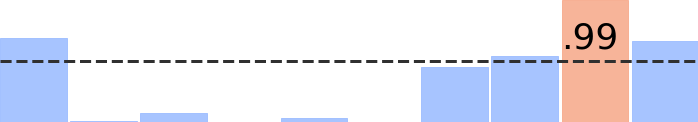}}\\
				& 23 & \cellcolor{Gray} I do enjoy playing bingo, if you'd like to join me for a game tomorrow night at church you're welcome to --- \textbf{Harry Sultenfuss, My Girl}& \cellcolor{Gray} \raisebox{-\totalheight\relax}{\includegraphics[width=0.15\textwidth]{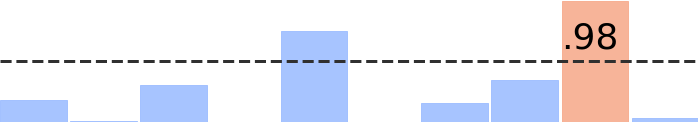}} \\
        & 24 &  Oh, I'm sure it's a lot of fun, 'cause the Incas did it, you know, and-and they-they-they were a million laughs --- \textbf{Alvy Singer, Annie Hall} &  \raisebox{-\totalheight\relax}{\includegraphics[width=0.15\textwidth]{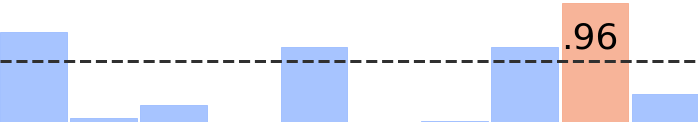}}\\
			\midrule 
        \multirow{3}{*}[-2.7mm]{\rotatebox[origin=c]{90}{Conflict}} & 25 & \cellcolor{Gray} Forgive me for askin', son, and I don't mean to belabor the obvious, but why is it that you've got your head so far up your own ass? --- \textbf{Gus Moran, Basic Instinct}  & \cellcolor{Gray} \raisebox{-\totalheight\relax}{\includegraphics[width=0.15\textwidth]{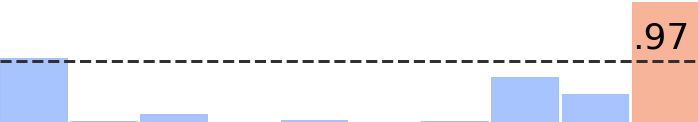}}  \\ 
				& 26 & If you're lying to me you poor excuse for a human being, I'm gonna blow your brains all over this car --- \textbf{Seamus O'Rourke, Ronin} & \raisebox{-\totalheight\relax}{\includegraphics[width=0.15\textwidth]{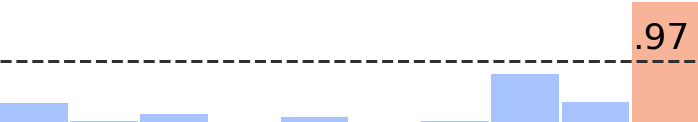}} \\ 
				& 27 &  \cellcolor{Gray} I couldn't give a sh*t if you believe me or not, and frankly I'm too tired to prove it to you --- \textbf{Evan Treborn, The Butterfly Effect}  & \cellcolor{Gray} \raisebox{-\totalheight\relax}{\includegraphics[width=0.15\textwidth]{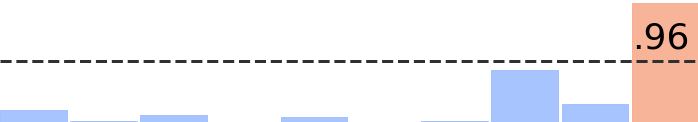}} \\ 
        \midrule 

    \end{tabular}
    \caption{The social dimensions in movie scripts. The three quotes with highest confidence score for each dimension are reported. For each quote, on the right, we report the histogram of the classifier confidence scores for all dimensions, and a horizontal line that marks a level of confidence of 0.5.}
    \label{tab:movie_quotes}
\end{table*}

\subsection{Predicting community outcomes}\label{sec:geostudy}

We saw that the 10 dimensions can be captured from conversations between pairs of people and reflect their relationships. We then tested whether the presence of those dimensions in conversations is associated with real-world outcomes at community-level. We expect to find such a connection because language is more than a mere communication medium. The words we use effectively reflect and change the reality around us~\cite{green2012pragmatics}, and the words that are used collectively by a community reveal the social processes associated to its thriving or decline. Since our Reddit data comprises of messages written by users that are geo-referenced at US State-level (\S\ref{sec:dataset:reddit}), we conducted a geographical analysis to study the relationship between the presence of the 10 dimensions and socio-economic outcomes. We set out to test three hypotheses:

\vspace{4pt}\noindent\emph{\textbf{H1}: Knowledge and education}. People with higher degrees have higher language proficiency~\cite{graham1987english} and are more likely to access and contribute to technical content online~\cite{glott2010wikipedia,thackeray2013correlates}. We hypothesize that US States with higher exchanges of knowledge are associated with higher education levels.

\vspace{4pt}\noindent\emph{\textbf{H2}: Knowledge and wealth}. Social networks in which knowledge is exchanged create innovation and technological advancements, which result into economic growth~\cite{florida2005cities,bettencourt2007invention}. We hypothesize that US States with higher exchanges of knowledge are also associated with higher per-capita income.

\vspace{4pt}\noindent\emph{\textbf{H3}: Trust, support, and suicides}. People affected by depression, especially those who have suicidal thoughts, do not tend to trust their peers~\cite{gilchrist2006barriers,shilubane2012psychosocial,cigularov2008prevents}, and seek social support in different contexts, often online~\cite{de2016discovering}. We therefore expect to find high levels of social support and reduced level of trust in States with high suicide rates.

\vspace{4pt}To verify these three hypotheses, we downloaded the 2017 American Community Survey statistics from the United States Census Bureau\cite{acs_survey}. The survey reports, for each State, the median household income and the proportion of residents with bachelor's degree or higher as a proxy for education levels. From the US Center of Disease Control~\cite{cdc_survey}, we downloaded the State-level suicide death rate calculated from the residents' death certificates. 

We ran our classifiers on every sentence of all the $\sim$160M posts and comments published by the $\sim$1M of Reddit users for which we estimated their State of residence. Similar to the analysis of Enron emails, we marked each text with dimensions $d$ whenever the confidence of model $M_d$ exceeded the threshold of 0.95 for at least one sentence in the text. Last, we estimated the prevalence of a dimension $d$ in a State as the number of posts labeled with $d$ normalized by the total number of posts in that State. 

\begin{figure}[t!]
    \centering
		\includegraphics[width=0.33\linewidth]{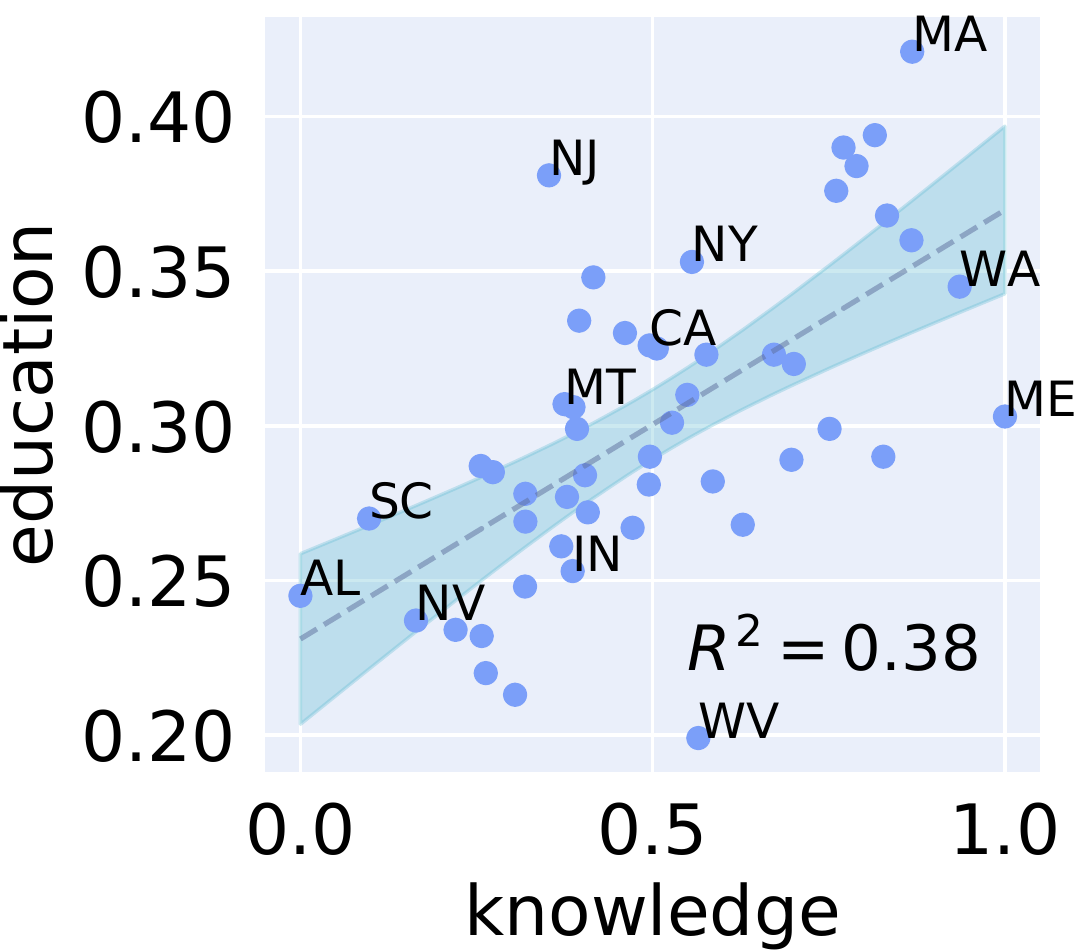}
		\includegraphics[width=0.32\linewidth]{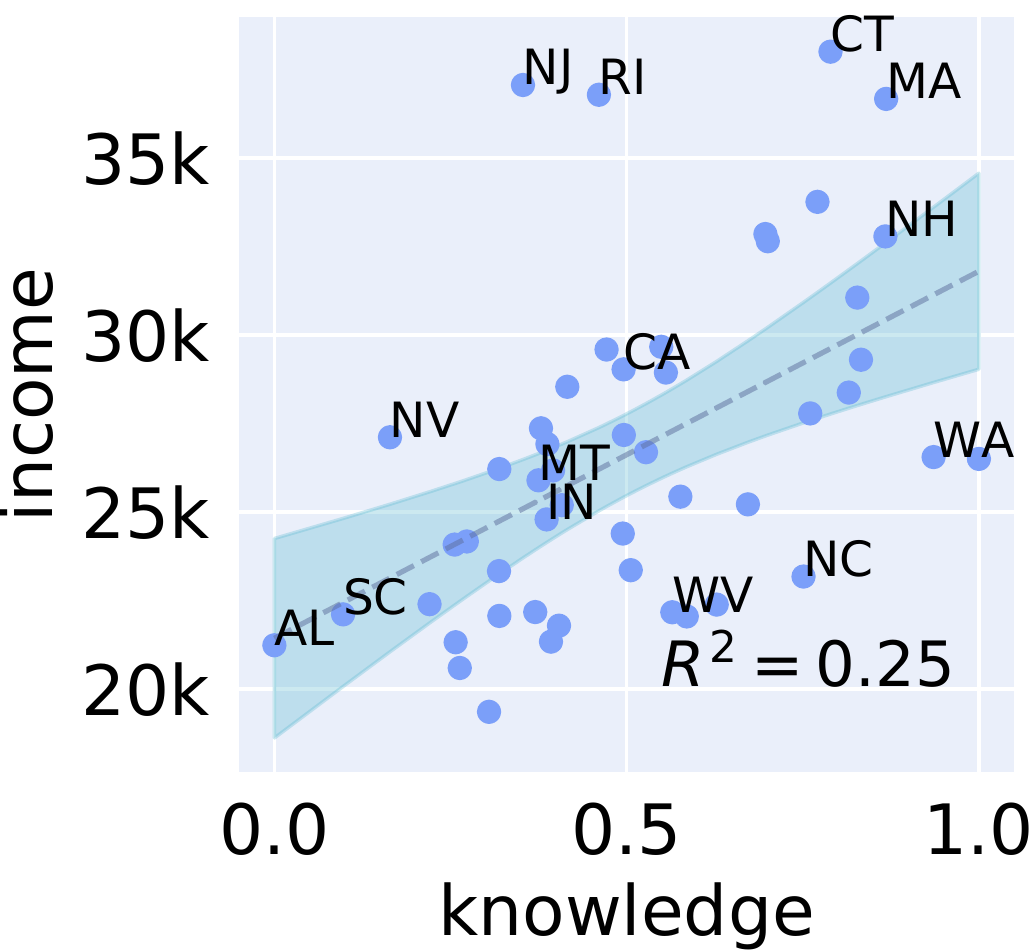}
    \includegraphics[width=0.31\linewidth]{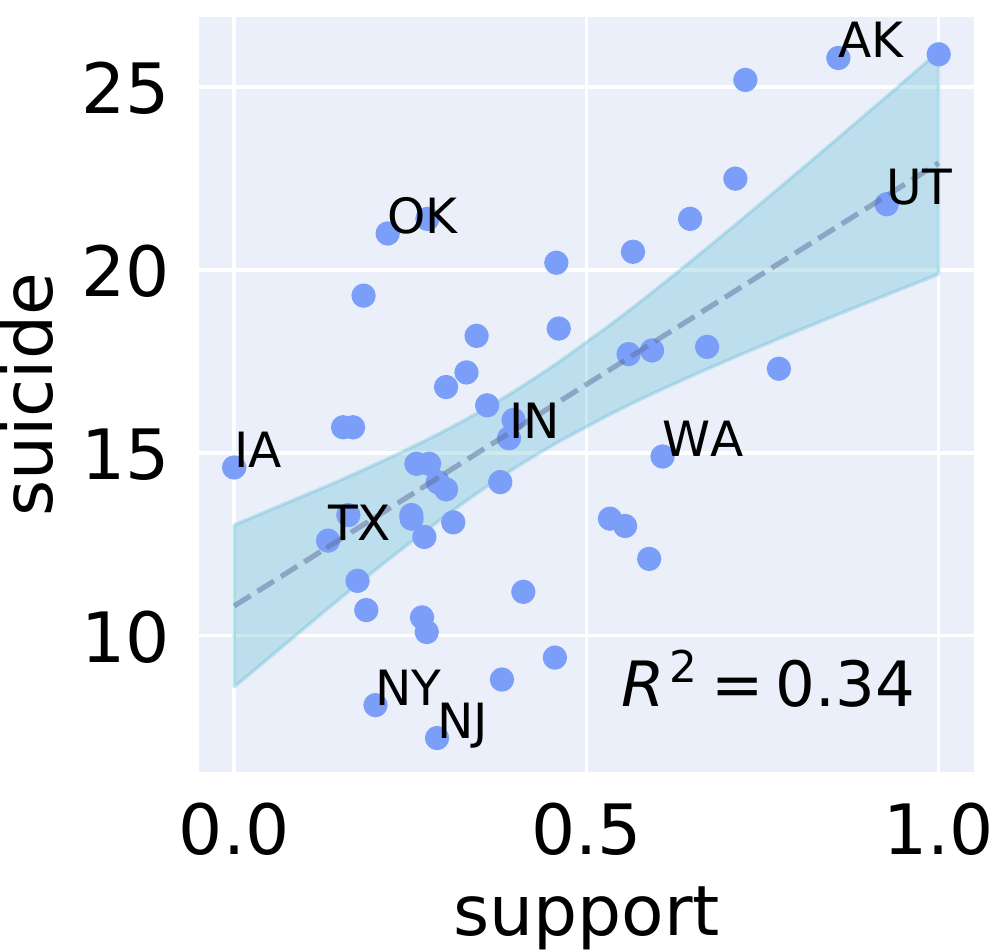}
    \caption{Linear relationships between each US-State outcome variable (education, income, suicide rate) and its most predictive social dimension (min-max normalized). Plots are annotated with a few representative US States.}
    \label{fig:geo_correlations}
		\vspace{-10pt}
\end{figure}

We ran a linear regression to estimate each of the census indicators from the State-level prevalence of the 10 dimensions. As a control factor, we added population density, which is associated with a number of socio-economic outcomes~\cite{bettencourt2013origins}. Overall, our hypotheses were confirmed (Table~\ref{tab:regression}). \emph{Knowledge} is the strongest significant predictor of education levels and income. Presence of \emph{support} and absence of \emph{trust} are the two most important predictors of suicide rates. As expected, population density alone is a good proxy for all the outcomes (urban areas are richer and more educated, with fewer cases of suicide). Yet, adding the conversational dimensions to the density-only baseline yields an absolute $R_{adj}^2$ increase between 0.25 to 0.52; with all the factors combined, all $R_{adj}^2$ exceed 0.7. Figure~\ref{fig:geo_correlations} displays the linear relationship between the outcome variables and the strongest predictors in the three regressions. 

A few other significant predictors emerge beyond what we hypothesized. States with higher education exhibit lower levels of \emph{conflict}. This is consistent with studies that found that hate speech is fueled by low education levels~\cite{gagliardone2015countering}. Wealth is associated with a reduced number of expressions that point out similarities between points of view, which might be a sign of communities that are structurally and culturally diverse~\cite{cummings2004work,lee2010knowledge}. Suicide rates are higher in States with fewer expressions of \emph{identity}, in line with previous studies that found an association between lack of sense of belonging and risk of depression-related suicides among young people~\cite{proctor1994risk}.

{\def\arraystretch{1.50}
\begin{table}[t!]
\footnotesize
\begin{center}
\begin{tabular}{lp{9mm}p{6mm}|p{9mm}p{6mm}|p{9mm}p{6mm}}
& \multicolumn{2}{c}{\textbf{Education}} & \multicolumn{2}{c}{\textbf{Income}} & \multicolumn{2}{c}{\textbf{Suicides}}\\
	& $\beta$	& SE & $\beta$ & SE & $\beta$ &	SE\\
\hline
                intercept	    &  .111          & .009 &  .233         & .099 & .228         & .109 \\
\rowcolor{Gray} Knowledge     &  \textbf{.554}$^{***}$  & .172 &  \textbf{1.140}$^{***}$ & .192 & .219          & .211   \\
                Power         &  .187          & .159 & -.209         & .177 & .004          & .195   \\
\rowcolor{Gray} Status        & -.217          & .199 &  .150         & .222 & .054          & .244   \\
                Trust         &  .309          & .205 & -.050         & .223 & \textbf{-.768}$^{***}$ & .251 \\
\rowcolor{Gray} Support       &  .278          & .238 &  .134         & .099 & \textbf{1.103}$^{***}$  & .291   \\
                Romance       & -.247          & .118 & -.182         & .133 & -.044         & .145  \\
\rowcolor{Gray} Similarity    & -.496          & .191 & \textbf{-.597}$^{***}$ & .214 & -.113         & .234  \\
                Identity      &  \textbf{.224}$^{*}$    & .126 & -.053         & .141 & \textbf{-.333}$^{**}$  & .154  \\
\rowcolor{Gray} Fun           &  .191          & .000  & -.127	       & .169 & .027          & .185   \\
                Conflict      & \textbf{-.300}$^{**}$   & .115 & -.211         & .127 & \textbf{.280}$^{*}$    & .141   \\
\rowcolor{Gray} Pop. density  &  \textbf{.433}$^{***}$  & .080 &  \textbf{.731}$^{***}$ & .090 & \textbf{-.614}$^{***}$ & .098  \\
\hline
$R_{adj}^2$ & \multicolumn{2}{r}{\textbf{.782} (+.522)} & \multicolumn{2}{r}{\textbf{0.774} (+.334)} & \multicolumn{2}{r}{\textbf{.707} (+.253)} \\
Durbin-Watson & \multicolumn{2}{r}{2.202} & \multicolumn{2}{r}{2.134} & \multicolumn{2}{r}{2.390} \\
\Xhline{2\arrayrulewidth}
\hline
\end{tabular}
\end{center}
\caption{Linear regressions that predict real-world outcomes (education, income, suicide rate) at US-State level from the presence of the 10 dimensions in the conversations among Reddit users residing in those States. Population density is added as a control variable. Absolute $R_{adj}^2$ increments of the full models over the density-only models are reported in parenthesis.}
\label{tab:regression}
\vspace{-10pt}
\end{table}
}

\section{Conclusion}\label{sec:discussion}

\subsection{Results and implications}

Starting from a unified theory that identifies the fundamental building blocks of social interactions, we collected data to associate these building blocks with verbal expressions, and we trained a deep-learning classifier to detect such expressions from potentially any text. Our tests obtained high prediction performances, showed that our tool correctly qualified the coexistence of different social dimensions in individual sentences and ascertained that the presence of certain dimensions is predictive of real-worlds outcomes.

From the theoretical standpoint, our work contributes to the understanding of how some of the fundamental sociological elements that define human relationships are reflected in the use of language. In particular, we discovered that all the 10 dimensions are represented abundantly in everyday conversations (albeit not equally), and that the way they are expressed can be learned even from a small number of examples. In practice, the data we collected and the classifiers we built could contribute to creating new text analytics tools for social networking sites. In particular, we believe that the dynamics of a number of processes mediated by social networks (including diffusion, polarization, link creation) could be re-interpreted with our application of the 10 dimensional model to conversation networks. To aid this process, we made our code and crowdsourced data available\footnote{\url{https://social-dynamics.net/projects/social_dimensions}} and encourage researchers to experiment with it, while considering the limitations we cover next.

\subsection{Limitations}

Our approach has limitations that future work will need to address. 

\vspace{4pt}\noindent\textbf{Data biases.} The data sources we used suffer from a number of biases. Our classifiers are trained on a restricted datasets from a single source (Reddit), made of texts posted by US residents, and labeled by annotators from English-speaking countries. As a result, some dimensions were underrepresented in the labeled data. A larger data collection with reduced socio-demographic, cultural, and linguistic biases is in order. We focused on phrases containing $1^{st}$ or $2^{nd}$ person pronouns and considered online conversations only; we did not test our tool on conversations happening offline.

\vspace{4pt}\noindent\textbf{Models.} Our models do not take into account important aspects of social interactions. First, they do not account for directionality. For example, a sentence classified as \emph{support} could either contain expressions of social support that the speaker is giving to others as well as the acknowledgment that others have provided support to the speaker. Second, we performed training focusing only on the sentences labeled by annotators, and not on the surrounding context. As a result, our models might fail to grasp the broader context around a phrase (e.g., Table~\ref{tab:movie_quotes}, line 7), which, for example, resulted in their inability to detect sarcasm (e.g., Table~\ref{tab:movie_quotes}, line 24).

\vspace{4pt}\noindent\textbf{Exhaustiveness of the 10 dimensions.} The theoretical model we operationalized is not meant to exhaustively map all the possible elements that define social interactions. Yet, the 10 dimensions summarize key concepts that have been extensively studied over decades in social and psychological sciences. Therefore, our analysis is comprehensive in that it includes the most frequent dynamics of interpersonal exchange. However, one might wonder why roughly 40\% of text samples could not be clearly labeled with any dimensions by the annotators (\S\ref{sec:results:crowdsourcing}). To investigate this aspect further, we manually inspected a sample of those instances. We found that, except a few instances of spam-like messages and false negatives, most sentences contained personal opinions on a matter (e.g., \emph{``My concern with this scenario is that she assumes that you would be into it.''}) or trivia (e.g., \emph{``My chinchilla attacks the vacuum the same way your rabbit attacks the broom''}). These are, to some extent, soft expressions of knowledge exchange or social support. In short, not all conversations convey a meaningful and clearly identifiable social meaning; a good part of it is generic chatter. Although we did not find any striking evidence that would point towards a need to revise or expand the underlying theoretical model, we still believe that further investigation across multiple datasets and scenarios is required. In conclusion, the ten dimensions might not be orthogonal and exhaustive representations of conversational language, yet we found that they express a very high descriptive power.

\section*{Acknowledgments}

We thank J\'er\'emie Rappaz, Eva Sharma, Tobias Kauer, Sebastian Deri, Miriam Redi, and Rossano Schifanella for their role in the creation of tinghy.org. We thank Daniel Romero and David Jurgens for their useful feedback on the paper draft.

\bibliographystyle{ACM-Reference-Format}

\end{document}